\documentclass[longauth,traditabstract]{aa} 
\usepackage{amssymb}
\usepackage{graphicx}
\usepackage{color}
\usepackage{subfigure}
\usepackage{natbib}
\bibpunct{(}{)}{;}{a}{}{,} 
\newcommand{\nc}{\newcommand}
\nc{\lsun}{\ensuremath{\mathrm{L}_\odot}}
\nc{\msun}{\ensuremath{\mathrm{M}_\odot}}
\nc{\tex}{\ensuremath{\mathrm{T}_{\rm ex}}}
\nc{\cthree}{C$_3$}
\nc{\cthreehtwo}{$c$-C$_3$H$_2$}
\nc{\kms}{\mbox{km\,s$^{-1}$}}
\nc{\Kkms}{\mbox{K\,km\,s$^{-1}$}}
\nc\micron{\mbox{$\mu$m}}
\nc{\Trot}{$T_{\rm rot}$}%
\nc{\Ntot}{$N(C_3)$}%
\nc{\Tc}{$T_{\rm c}$}%
\nc{\Tdust}{$T_{\rm dust}$}%
\nc{\Tex}{$T_{\rm ex}$}%
\nc{\Tkin}{$T_{\rm kin}$}%
\nc{\Tmax}{$T_{\rm max}$}%
\nc{\cmcub}{\mbox{cm$^{-3}$}}
\nc{\cmsq}{\mbox{cm$^{-2}$}}
\newcommand{\HII}{H {\sc ii}}

\newcommand\arcdeg{\mbox{$^\circ$}}%

\newcommand{\crrl}{C91$\alpha$}

\newcommand{\CII}{[C {\sc ii}]}
\newcommand{\OI}{[O {\sc i}]}

\def\Lsun{\,{\rm L$_{\odot}$}}

\nc{\thCO}{$^{13}$CO}
\nc{\thCII}{[$^{13}$C {\sc ii}]}

\begin{document}

\title{High Spectral and Spatial Resolution Observations of the PDR Emission in the NGC2023 Reflection Nebula with SOFIA and APEX \thanks{The NASA/DLR
Stratospheric Observatory for Infrared Astronomy (SOFIA) is jointly
operated by the Universities Space Research Association, Inc. (USRA), under
NASA contract NAS2-97001, and the Deutsches SOFIA Institut (DSI) under DLR
contract 50 OK 0901 to the University of Stuttgart.}~\thanks{
The Atacama Pathfinder Experiment
(APEX) is a collaboration between the Max-Planck-Institut f\"ur
Radioastronomie, the European Southern Observatory, and the Onsala Space
Observatory.}~\thanks{{\it Herschel} is an ESA space observatory with science instruments provided by European-led Principal Investigator consortia and with important participation from NASA.}}
   \author{G. Sandell\inst{1}
 \and  B.~Mookerjea\inst{2}
  \and  R. G\"usten \inst{3}
  \and  M. A. Requena-Torres \inst{3}
   \and  D. Riquelme \inst{3}
     \and  Y. Okada \inst{4}
}

\institute{SOFIA-USRA, NASA Ames Research Center, MS 232-12, Building N232, Rm.
146, PO Box 1, Moffett Field, CA 94035-0001,
USA \email{Goran.H.Sandell@nasa.gov} \label{sofia}
\and
Tata Institute of Fundamental Research, Homi Bhabha Road,
Mumbai 400005, India \email{bhaswati@tifr.res.in}\label{tifr} 
\and
Max Planck Institut f\"ur Radioastronomie, Auf dem H\"ugel 69, 53121 Bonn, Germany
\and
I. Physikalisches Institut der Universit\"at zu K\"oln, Z\"ulpicher Stra{\ss}e 77, 50937 K\"oln, Germany
}

\date{Accepted  April 13, 2015}
\abstract
{We have mapped the NGC 2023 reflection nebula in \CII\ and CO(11--10)
with the heterodyne receiver GREAT on SOFIA and obtained slightly smaller
maps in $^{13}$CO(3--2), CO(3--2),  CO(4--3), CO(6--5), and CO(7--6) with
APEX in Chile.  We use these data to probe the morphology, kinematics, and
physical conditions of the C~{\sc ii} region, which is ionized by FUV
radiation from the B2 star HD\,37903. The \CII\ emission traces an
ellipsoidal shell-like region at a position angle of  $\sim$
$-$50\arcdeg, and is surrounded by a hot molecular shell. In the southeast,
where the C  {\sc ii} region expands into a dense, clumpy molecular cloud
ridge, we see narrow and strong line emission from high-$J$ CO lines, which
comes from a thin, hot  molecular shell
surrounding the \CII\ emission.  The \CII\ lines are broader and show photo
evaporating gas flowing into the C~{\sc ii} region. Based on the strength
of the \thCII\ F=2--1 line, the \CII\ line appears to be somewhat optically
thick over most of the nebula with an optical depth of a few.  We model the
physical conditions of the surrounding molecular cloud and the PDR emission
using both RADEX and simple PDR models. The temperature of the CO emitting
PDR shell is $\sim$ 90  -- 120 K, with densities of 10$^5$ -- 10$^6$ \cmcub, as
deduced from RADEX modeling. Our PDR modeling indicates that the PDR layer
where \CII\ emission dominates has somewhat lower densities, 10$^4$ 
to a few times 10$^5$ \cmcub.
}


\titlerunning{The NGC2023 PDR}
\authorrunning{Sandell et al.}
\maketitle

\section{Introduction}

Located at a distance of 350\,pc in the L\,1630 molecular cloud NGC\,2023
is one of the most
well-studied reflection nebulae in the whole sky. This is where the
polycyclic aromatic hydrocarbon (PAH) molecules were first identified
in  the interstellar medium \citep{Sellgren84}. NGC\, 2023 has a
favorable geometry, which makes it an almost ideal laboratory for
studying the physical conditions in photon dominated regions (PDRs).
HD 37903, the B2 V star illuminating the reflection nebula, is on the
near side of the molecular cloud, while the shape of the
reflection nebula indicates that the cloud interface is seen largely
edge-on in the southeastern part of the nebula, where it expands 
into a dense molecular cloud ridge.

The PDR emission from NGC\, 2023 has been studied extensively.
\citet{Howe91} used a Fabry-Perot spectrometer on the Kuiper Airborne observatory 
(KAO) to image the nebula
in the 158 $\mu$m \CII\ line with a spatial  resolution of
55\arcsec, while
 \citet{Steiman97} observed the far-infrared fine structure emission  lines
of [OI] 63 and 145 $\mu$m,  \CII\  158 $\mu$m, and  [Si II] at 35 $\mu$m with
a cryogenic grating spectrometer. \citet{Jaffe90}, using ground based
telescopes,  studied the PDR region by observing the CO(2--1), CO(3--2),
CO(7--6), and C$^{18}$O(2--1) transitions and found that there was a bright
ridge to the southeast of HD\,37903, where the CO(7--6) emission coincides
with \CII\ emission and fluorescently excited H$_2$ emission. This ridge
was resolved into a lumpy filament  in the C91$\alpha$ recombination
line by \citet{Wyrowski00}, who used the Effelsberg 100 m telescope and 
the VLA with  an angular resolution of 11\arcsec. The nebula has been
imaged in vibrationally excited H$_2$ emission by \citet{Gatley87},
\citet{Field94,Field98}, \citet{McCartney99}, and \citet{Martini99},
revealing bright filamentary H$_2$ emission especially to the south and
southwest. \citet{Fleming10}, \citet{Sheffer11}, and \citet{Peeters12} used
the  Infrared Spectrograph (IRS) on the {\it Spitzer Space Telescope} to
observe pure rotational H$_2$ lines and PAH emission in several regions of
NGC\,2023 in order to compare the observed intensities with predictions
from PDR models. \citet{Mookerjea09} analyzed IRAC and MIPS images  from
the {\it Spitzer Space Telescope} as well as SCUBA images from the James
Clerk Maxwell Telescope (JCMT). They found that HD\,37903 is the most
massive member of a cluster with 20 -- 30 pre-main-sequence  (PMS) stars as
well as at least  two young stellar objects (YSOs), MM\,3 and
MM\,4, embedded in the southern and southeastern part of the reflection
nebula.

\section{Observations}

\subsection{\CII\ and CO(11--10) with SOFIA}

The \CII\ emission  was mapped with  the heterodyne receiver GREAT
\footnote{GREAT is a development by the MPI  f\"ur Radioastronomie
(Principal Investigator: R. G\"usten) and the KOSMA/ Universit\"at zu
K\"oln, in cooperation with the MPI  f\"ur Sonnensystemforschung and the
DLR Institut f\"ur Planetenforschung.}\citep{Heyminck12}  onboard the
Stratospheric Observatory for Infrared Astronomy (SOFIA) during three cycle
1 flights: November 1, 2013, February 4 and February 5, 2014. On all
flights the GREAT L1 channel was tuned to CO(11--10) in the lower sideband.
Due to a local oscillator failure the L2 mixer (targeting \CII{}) was
restored to its basic science configuration for the  February flights,
resulting in somewhat lower beam efficiencies due to changes  in the optics
(see below). The flight on November 1 2013 provided excellent observing
conditions at 13.1 km with system temperatures of $\sim$ 2450 K for the L1
mixer and $\sim$ 3,000 K for L2. In February the L1 system temperatures
were about the same as in November  (2500 -- 2600 K), while the system
temperatures for L2 were about 3,000 K. The beam efficiency, $\eta_{mb}$,
for the November 2013 flight series was measured to be 0.67 for both
channels, while it was 0.62 for L2 during the February flights.
Table~\ref{tbl-1} lists rest frequencies, half power beamwidths (HPBWs) and
beam efficiencies for all receiver configurations. All beam efficiencies
are based on observations of Jupiter.

The whole nebula was mapped in total-power on-the-fly (OTF) mode with 1
s integration time at each dump, and 7.5\arcsec\ step size. The reference
position, at +400\arcsec, $-$90\arcsec\ offset from HD\,37903, has some
faint (a few K) \CII\ emission at $\sim$ 8 \kms,  which was corrected for
in the post processing.  The map was  built up by doing a number of small,
overlapping maps. The final map size is $\sim$ 6.6\arcmin\ in RA and $\sim$
6.8\arcmin\ in Dec, covering most of the optically visible reflection
nebula. Long (5 --10 min) integrations were obtained at three positions:
(0\arcsec,0\arcsec{}), ($+$30\arcsec,$+$15\arcsec{}), and
($+$45\arcsec,$-$45\arcsec{}); all measured relative to HD\,37903
($\alpha$(2000.0) =$5^h41^m38.^s388$, $\delta$(2000.0) = $-$02\arcdeg\
15\arcmin\ 32.5\arcsec{}). The spectra had very good baseline stability and
in  most cases nothing higher than a 2nd order polynomial baseline was
removed. The final map cubes were created using CLASS\footnote{CLASS is
part of the Grenoble Image and Line Data Analysis Software (GILDAS), which
is provided and actively developed by IRAM, and is available at
http://www.iram.fr/IRAMFR/GILDAS} and were re-gridded to 6\arcsec\  spatial
grid for both \CII\ and CO(11--10). Maps were created  with several
different velocity resolution by resampling the spectra to either 0.25, 0.5
and 1 \kms, in order to get a better idea of the morphology and velocity
structure of the emission. The rms noise per resolution element  is $\sim$
1.6 K for the C(11--10) map and 2.5 K for \CII,  both for a velocity
resolution of 0.25 \kms.
 
\subsection{APEX observations of CO(3--2), \thCO(3--2),  CO(4--3), CO(6--5), \& CO(7--6).}

NGC\,2023 was observed  on December 1, 2013 using the FLASH$^+$
receiver on the 12~m Atacama Pathfinder EXperiment (APEX) telescope,
located at Llano de Chajnantor in the Atacama desert of Chile
\citep{Gusten06}. FLASH$^+$ is a dual channel heterodyne SIS receiver
operating simultaneously  on orthogonal polarizations in the 345 GHz and in
the 460  GHz atmospheric windows \citep{Klein14}. Both bands employ
state-of-the-art sideband separating SIS mixers. The mixers provide large
tuning ranges enabling simultaneous observations of  $^{13}$CO and
$^{12}$CO(3--2) in the 345 GHz window and $^{12}$CO(4--3) in the 460 GHz
window. The backends are Fast Fourier Transform Spectrometers
\citep{Klein12} with a total bandwidth of 4 GHz and 76.3 kHz  frequency
resolution. The observations were carried out in good observing conditions
resulting in system temperatures of 150 -- 180 K for $^{13}$CO and
$^{12}$CO(3--2) and 340 -- 380 K for $^{12}$CO(4--3), respectively. A
region of 5.3\arcmin\ $\times$ 6.3\arcmin\ was mapped in on-the fly mode
with a  spacing of 6\arcsec\ in RA and Dec and all scans were repeated
once. The maps were created  using CLASS  and calibrated in T$_{mb}$ using
main beam efficiencies measured on Jupiter, appropriate for a somewhat
extended source  (see Table~\ref{tbl-1}). The quality of the individual
spectra were excellent and only a linear baseline was  removed from the
data. Map cubes were created with the same velocity resolution and on the
same spatial grid as for the SOFIA GREAT observations.  For a velocity
resolution of 0.25 \kms, the rms noise in the $^{13}$ CO(3--2), CO (3--2)
and CO(4--3) are  about 0.4,  0.3, and 0.6 K, respectively.

On October 31, 2014 the same area was mapped in CO(6--5) and CO(7--6)
using CHAMP$^+$ on APEX. CHAMP$^+$ is a dual channel 7-pixel hexagonal
SIS-mixer  heterodyne array with a central pixel. It works simultaneously
in  the 450 and 350 $\mu$m atmospheric windows \citep{Kasemann06}. The
weather conditions were excellent with precipitable water vapor  (PWV) in
the range 0.54 to 0.35 mm. The system temperatures for CO(6--5) were on the
average $\sim$ 1200~K.  At the very end of the observing run the system
temperatures went up to $\sim$ 1600 K. For CO(7--6) the system temperatures
were $\sim$ 4000~K for most mixers, except for two mixers which were about
2000~K higher. Because the area covered is quite large, the field was
divided into 10 overlapping subfields. Observations were done in OTF mode,
sampling data every 4\arcsec. The reference position was at $-$1385\arcsec,
$-$274\arcsec\ relative to HD\,37903 and has been verified to be free of CO
emission. Each field was covered twice, first scanning in RA and then in
DEC, resulting in maps with fairly uniform depth except at the outskirts of
the coadded map where the noise level is naturally higher. All spectra are
calibrated in T$_{mb}$ using main beam efficiencies measured on Jupiter
(see  Table~\ref{tbl-1}).  The pointing was checked on the nearby  $\alpha$
Ori in the CO(6--5) line and is good to 2.5\arcsec. For a velocity
resolution of 0.25 \kms, the rms levels are $\sim$ 1.8 K and 3 -- 3.3 K for
CO(6--5) and CO(7--6),  respectively.

\begin{table}[h]
\begin{center}
\caption{Observing setup\label{tbl-1}} 
{\scriptsize
\begin{tabular}{lllcl}
\hline\hline

Receiver & Molecular transition &  Frequency   & $\theta_{FWHM}$ & $\eta_{mb}$ \\
 & & [GHz] & [~\arcsec{}~]  &  \\
\hline
GREAT L1 & CO(11--10) &  1267.01449 & 23.0 & 0.67\\
GREAT L2 & \CII\ $^{2}P_{3/2}\rightarrow^{2}P_{1/2}$	 & 1900.53690 & 15.3 & 0.67$^a$\\
GREAT L2 & \thCII\  F = 2--1	 & 1900.46610 & 15.3 & 0.67$^a$\\
FLASH$^+$ & $^{13}$CO(3--2) & 330.587965 & 18.5 & 0.68 \\
FLASH$^+$ & CO(3--2) & 345.795990 & 17.7 & 0.68 \\
FLASH$^+$ & CO(4--3)   & 461.040768 & 13.3 & 0.58\\
CHAMP$^+$ & CO(6--5) & 691.473076 & 9.1 & 0.49 \\
CHAMP$^+$ & CO(7--6) & 806.651806 & 7.7 & 0.48 \\
\hline
\hline
\end{tabular}}
\end{center}
{\noindent $^a$ $\eta_{mb}$ = 0.62 for the February 2014 flights.}

\end{table}


\section{Results}

\subsection{Overall morphology}
\label{sect:morphology}

\begin{figure*}[t]
\begin{center}
\includegraphics[width=\textwidth]{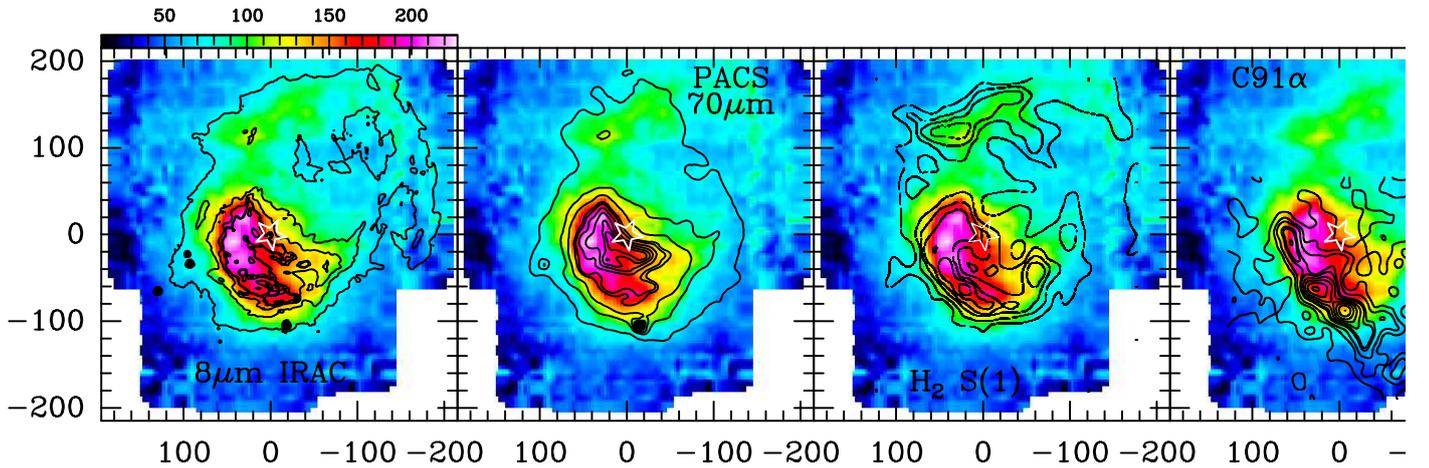}
\caption{Integrated intensity image of the \CII\ line at 158\,\micron\
in color. Contours correspond to the tracers mentioned in the individual
panels. The peak intensities (and contour levels as \% of peak) are
8\,\micron\ continuum image: 5306\,MJy\,sr$^{-1}$  (2\% to 10\% (2.5\%
step), 10 to 50\% (5\% step), 50 to 100\% (20\% step)), PACS 70$\mu$m
1.32\,10$^4$\,MJy/sr (10\%, 20\%, 30\%, 45\%, 50\% to 100\% (10\% step)),
\crrl: 7.3~10$^{-3}$\,Jy/beam (5\%, 20\%, 30\%, 45\%, 50\% to 100\% in
steps of 10\%), H$_2$ S(1):  7.3~10$^{-20}$\,W\,cm$^{-2}$/beam (10\% to
90\% in steps of 10\%). The \crrl\ image is from \citet{Wyrowski00}
and the H$_2$ S(1) image is from \citet{Gatley87}. The offsets in all
figures are measured in arcseconds relative to HD\,37903, which is marked
by a star symbol. 
\label{fig_intplots}}
\end{center}
\end{figure*}

\begin{figure*}[t]
\begin{center}
\includegraphics[width=\textwidth]{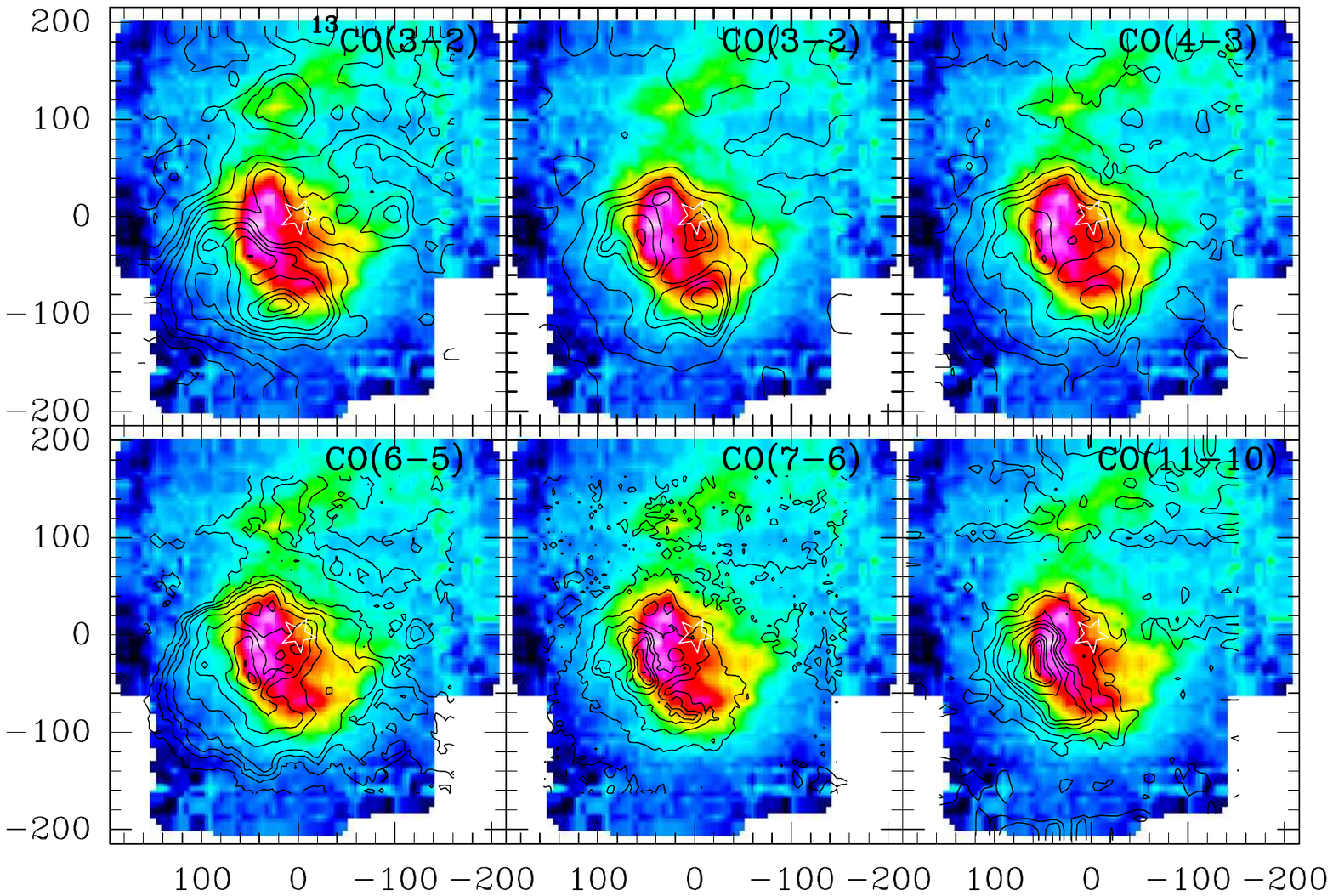}
\caption{Integrated intensity image of the \CII\ line at 158\,\micron\
in color.  Contours correspond to the tracers mentioned in the individual
panels. The contour levels are \thCO(3--2): (3--8), 
(10--13),(15--18)$\sigma$ with $\sigma = 4.2$\,K\kms. CO(3--2): (3--8), 10,
15, 20, 25, 30, 32, 35, 38, 40 and 42 times $\sigma = 5.1$\,K\kms.
CO(4--3): (3--8), 10, 12, 15, 18, 21, 24 times $\sigma = 8.6$\,K\kms.
CO(6--5): 3, 5.5, 9.5, 11.5, 13, 15, 20, 25, 30, 35, 40 times $\sigma =
5.4$\,K\kms. CO(7--6): 4, 6, 11, 15, 18, 20, 22, 24, 26 times $\sigma =
8.2$\,K\kms. CO(11--10): 3, 5, 8, 10, 12, 15, 20, 22, 24, 26 times $\sigma
= 1.3$\,K\kms.  HD\,37903 is marked by a star symbol.
\label{fig_intcoplots}}
\end{center}
\end{figure*}

The \CII\ emission appears to trace an ellipsoidal (egg-shaped) region
oriented approximately southeast to northwest  (position angle $\sim$
$-$50\arcdeg{}) (Fig.\,\ref{fig_intplots}). The emission is very strong in
the southeast where the \CII\ region expands into a dense molecular cloud
ridge, which was mapped by  \citet{Wyrowski00} in HCN(1--0) and
HCO$^+$(1--0). In the northwest the \CII\ emission  is more diffuse and
extended. In the southeastern quadrant of the nebula the emission is very
bright with main beam brightness temperatures between 30 - 75 K. The \CII\
emission shows the same morphology as the PAH emission (at 8 $\mu$m), PACS
70 $\mu$m emission, and vibrationally excited H$_2$ emission, while
C91$\alpha$ is much more localized and only seen in the southeastern part
of the nebula where the PAH emission is very strong
(Fig.\,\ref{fig_intplots}). Both PAHs and vibrationally excited H$_2$ are
excited by FUV radiation. The 70 $\mu$m emission is dominated by emission
from small dust grains, which are also  heated  by photoelectric heating
from the FUV radiation. Therefore it is clear that  the  excitation of the
\CII\ in this area is also dominated by the FUV field from HD\,37903, the
B2 star illuminating the nebula.

\begin{figure}[h]
\begin{center}
\includegraphics[width=0.49\textwidth]{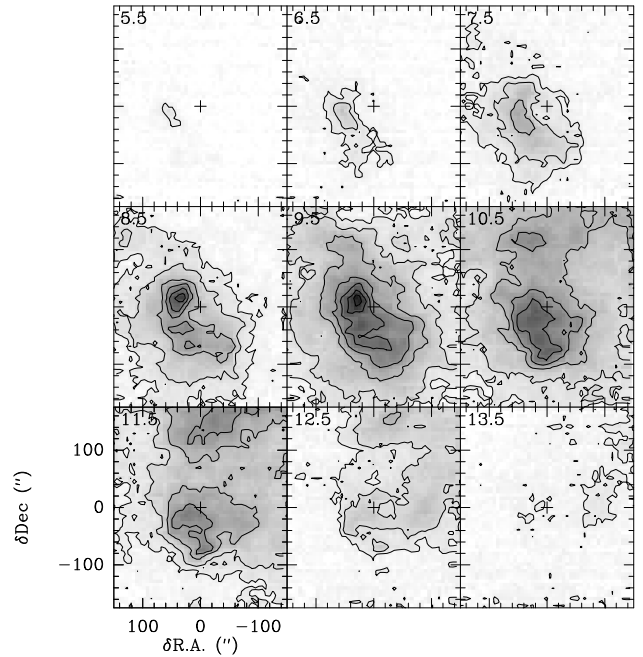}
\caption{Velocity-channel maps for \CII\  in gray scale overlaid with contours at 5 and 10\,K, and from there
 to 70\,K in steps of 10\,K. The `+' shows the position of HD\,37903.
\label{fig_cpluschanmap}}
\end{center}
\end{figure}

Although low-$J$ CO emission is detected over the entire mapped region
(Fig. \ref{fig_intcoplots}), the strongest emission is to the south and
southeast of the nebula. The $^{13}$CO(3--2) emission, which traces the
column density rather than the temperature of the gas, shows the dense
ridge to the southeast. It also has a secondary peak coinciding with the
northern \CII\ emission peak at ($\sim$ 20\arcsec, 110\arcsec, ) indicating
that the \CII\ emission is brighter where the FUV radiation illuminates a 
denser region of the surrounding cloud (Fig.\,\ref{fig_intcoplots}). The
CO(11--10)  emission is only visible in the southeastern quadrant of the
map, where the C {\sc II} region is bounded by the dense molecular cloud
located southeast and south of HD\,37903.  The vibrationally excited H$_2$
emission imaged with high spatial resolution indicates that the interface
between the ionized inner region and the surrounding dense molecular cloud
is far from smooth \citep{Field94,Field98,McCartney99}, with lumpy ridges
and bright filamentary structures. The same is true for the PAH emission
\citep{Fleming10,Sheffer11,Peeters12}. Our maps of \CII\ and high-$J$ CO
lines  (Fig.\,\ref{fig_intcoplots} \& \ref{fig_cpluschanmap}) look
smoother, mostly because of insufficient spatial resolution to see such
details. The \CII{}-map (Fig.\,\ref{fig_intplots}) looks very similar to
one of the early vibrationally excited H$_2$ 2.12 $\mu$m maps
\citep{Gatley87}, imaged with 20\arcsec{}-resolution.

\begin{figure}[h]
\begin{center}

\includegraphics[width=0.49\textwidth]{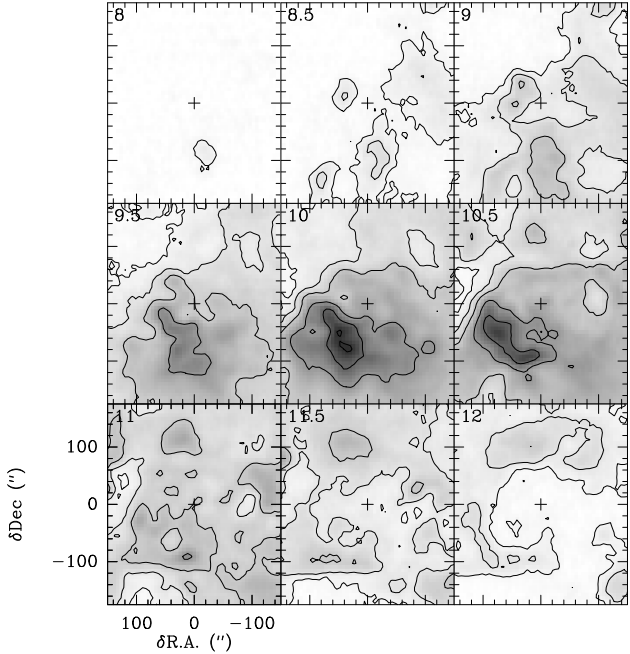}
\caption{Velocity-channel map for $^{13}$CO(3--2) in grey scale overlaid with contours at 2, 5, and
10 K, and from there to 50\,K in steps of 10\,K. The `+' shows the position of HD\,37903.
\label{fig_13co32chanmap}}
\end{center}
\end{figure}

Inspection of  the \CII\ and   \thCO{}(3--2) channel maps
(Fig.\,\ref{fig_cpluschanmap} \& \ref{fig_13co32chanmap}) show that there
is a clear velocity gradient from south to north.  This velocity gradient
is very obvious in the three position velocity plots (Figure
\ref{fig_pvplots}), which cut through HD\,37903 at three different position
angles: east to west, southeast to northwest and north to south. The
systemic velocities are $\sim$ 10 km~s$^{-1}$ in the south, $\sim$ 10.5
km~s$^{-1}$  in the southeast and east, while the velocities are   $\sim$
12  km~s$^{-1}$ in the north and northwest. The $^{13}$CO(3--2) channel
maps show a shell-like structure surrounding the whole nebula at
red-shifted velocities (Fig.\,\ref{fig_13co32chanmap}), indicative of an
expansion of the nebula.

The low-$J$ CO channel maps (Fig.\,\ref{fig_co32chanmap} \&
\ref{fig_co43chanmap}) show the same velocity gradient as seen in \CII\ and
$^{13}$CO, but in addition they show blue- and red-shifted emission in the
southern part of the image, with the strongest high velocity emission near
or centered on the extreme Class I object MM\,3 \citep{Mookerjea09}, also
known as Sellgren D or MIR-63. MM\,3, which is a relatively strong
millimeter/sub-millimeter source \citep{Wyrowski00, Mookerjea09},
definitely drives a bipolar molecular outflow, but there appear to be
several other fainter outflows in the region, including a blue-shifted
outflow lobe from Sellgren C (MIR--62) coinciding with the Southern Ridge
(SR).  All outflows are associated with young Class II or Class I objects.
These will be discussed in more detail in  a separate paper (Sandell et al.
2015, in prep). There is  a red-shifted velocity feature in the
northwestern corner of our CO maps without any  associated blue-shifted
emission. This is most likely part of the surrounding molecular cloud,
which appears to be more red-shifted in the north and northwest.

\subsection{A thin, hot molecular shell?}
\label{sect:coshell}

\citet{Jaffe90},  who observed the PDR region in  NGC\,2023 in
CO(2--1), CO(3--2), CO(7--6), C$^{18}$O(2--1), and \CII, found that
southeast of HD\,37903 the CO(7--6) lines  peak in a bright ridge
coinciding with \CII\ and fluorescently excited H$_2$ emission, and is
surrounded by a dense shell of warm molecular gas. They concluded that this
emission must originate in a region where the density is $\sim$ 10$^5$
cm$^{-3}$ and where the kinetic temperature is $\geq$ 85 K. Since our
observations include completely sampled maps of the reflection nebula in
both low-$J$ CO lines (CO(3--2) and CO(4--3)), as well as in high-$J$ CO
lines (CO(6--5), CO(7--6) and CO(11--10)) and  \CII\ with  high spatial and
spectral resolution, we are uniquely positioned to  detect where the warm,
hot, and ionized gas resides in the nebula.

The star illuminating the reflection nebula, HD\,37903, is a B2 V star.
The low visual extinction toward the star suggests that the star lies on
the near side of the molecular cloud \citep{Howe91}. The strong PDR
emission  southeast of the star looks like a half shell and suggests that
here the expanding \CII\ shell is seen almost edge on.

In order to understand the geometry of the hot molecular shell, the
CO(11--10) emission is first examined. This emission  only probes hot gas
associated with the PDR/molecular cloud interface with negligible
contribution from the cold molecular cloud surrounding the reflection
nebula. The CO(11--10) emission, however, is only seen in the southeastern
quadrant of the map, the region with the strongest PDR emission. Since the
peak  brightness temperature of CO(11--10) is $<$ 22 K, which is much less
than the 85 K that \citet{Jaffe90} inferred from their CO(7--6)
observations, it must be largely optically thin or sub thermally excited.
Therefore it is a good tracer of the hot gas in the PDR. The CO(7--6) and
CO(6--5) emission is also dominated by the hot PDR emission with CO(6--5)
having T$_{mb}$ $\sim$ 85 K, and CO(7--6) $\sim$ 60 K. The hot PDR emission
is also seen in CO(4--3) and CO(3--2), but these  lines are dominated by
optically thick emission from the surrounding molecular cloud and are
strongly affected by self-absorption from colder foreground gas. This makes
it difficult or impossible for these lines to separate the PDR emission
from the emission originating in the surrounding  molecular cloud.

\begin{figure*}[!]
\begin{center}
\includegraphics[width=0.95\textwidth]{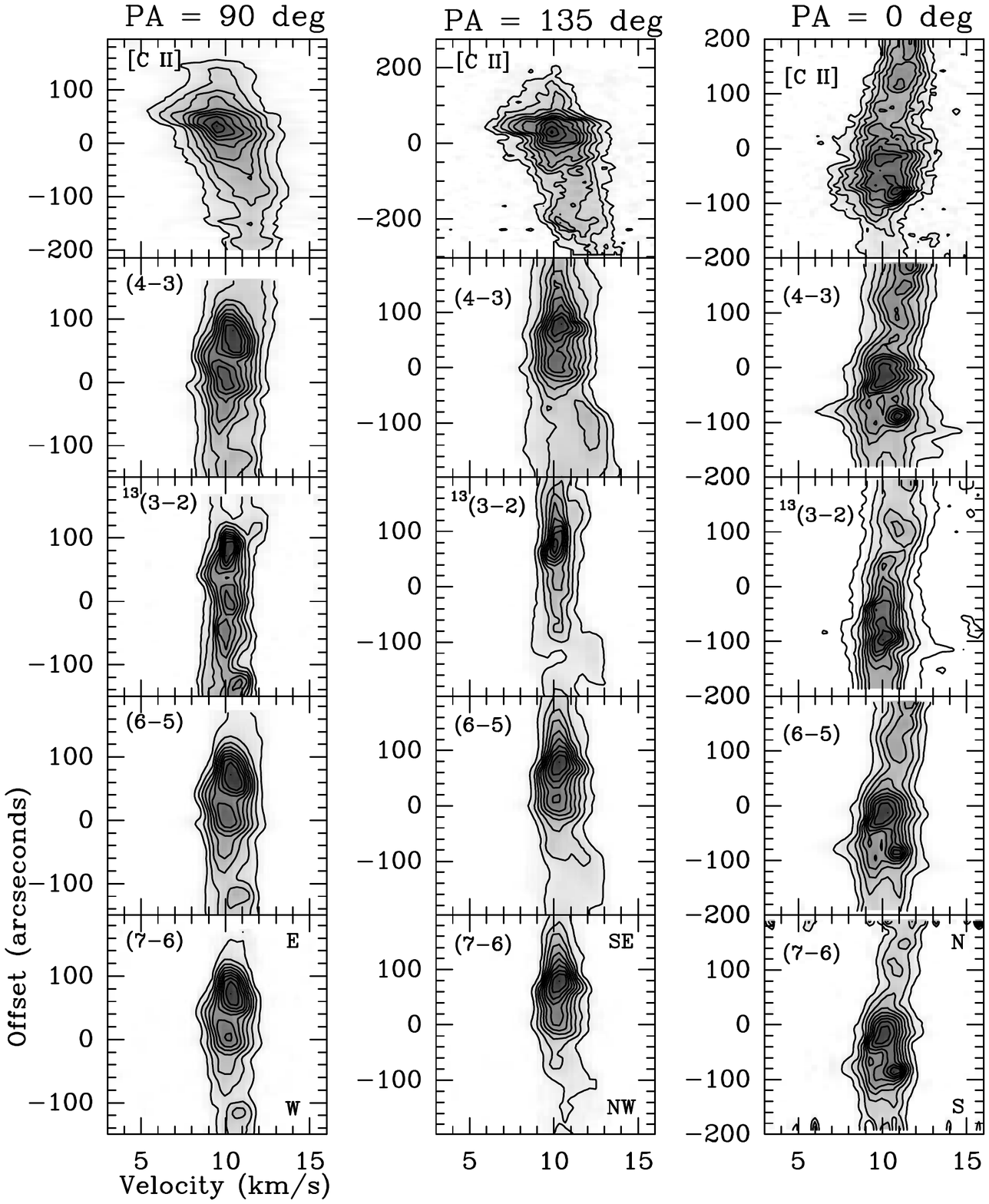}
\caption{Position-Velocity diagrams  for \CII, CO(4--3), $^{13}$CO(3--2), 
CO(6--5), and CO(7--6) along directions given by position angle, PA of 0, 90 and
135\arcdeg. The contours are at 10\% to 100\% (in steps of 10\%) of the
maximum T$_{\rm mb}$ of each plot. For each column we give the maximum
$T_{\rm mb}$ from top to bottom. For PA=90\arcdeg\ the values are  56.9, 72.8, 24.3, 69.5 and 54.9\,K. For
PA=135\arcdeg\ the values are 53.5, 80.5, 41.5, 78.2 and 62.3\,K. For
PA=0\arcdeg\ the values are 53.6, 64.3, 29.2, 56.6 and 43.3\,K. HD\,37903 is at 0\arcsec. 
The position angle is measured counterclockwise from north.
\label{fig_pvplots}}
\end{center}
\end{figure*}

\begin{figure}[h]
\begin{center}
\includegraphics[width=0.49\textwidth]{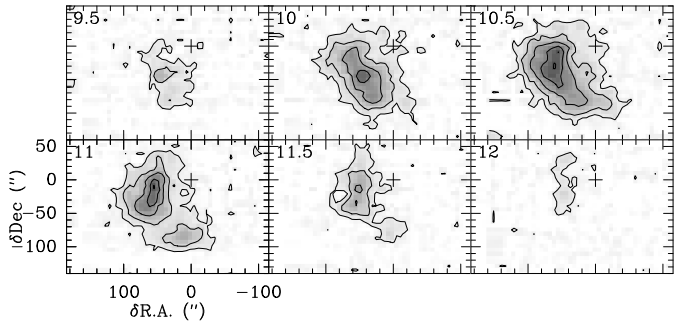}
\caption{Velocity-channel map for CO(11--10) with contours at  2.5 to 
30\,K in steps of 5\,K. The `+' shows the position of HD\,37903.
\label{fig_co11chanmap}}
\end{center}
\end{figure}

\begin{figure}[h]
\includegraphics[width=0.5\textwidth]{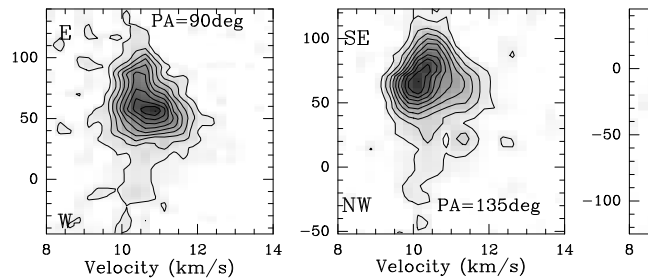}
\caption{Position-Velocity diagrams along selected directions for CO(11--10).
The contour levels for PA=90 and 135\arcdeg\ are 10\% to 100\% (10\%
step) of the maximum T$_{\rm mb}$ of 18.4 and 18.8\,K respectively. The
contour levels for PA=0\arcdeg\ are 20\% to 100\% (20\% step) of the
maximum T$_{\rm mb}$ of 9\,K.  HD\,37903 is at 0\arcsec. 
\label{fig_co11pvplots}}
\end{figure}

\begin{figure}[h]
\begin{center}
\includegraphics[width=0.49\textwidth]{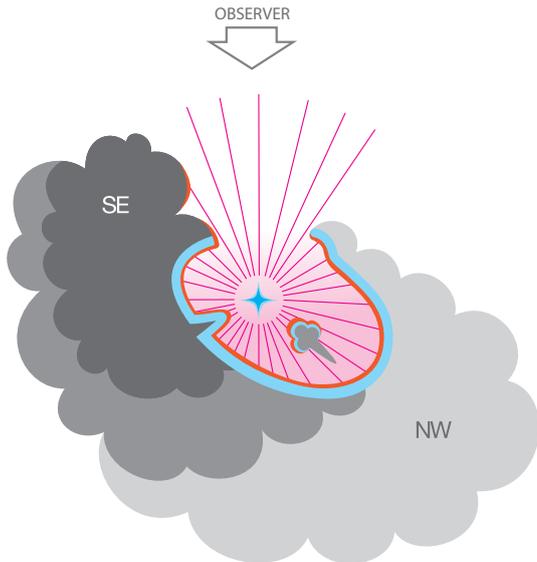}
\caption{Cartoon showing a simplified picture of hot, slowly expanding
PDR region surrounding HD\,37903. The surrounding cloud is shown in grey
with denser gas having a darker shading. The hot molecular shell is shown
in blue and the red shows the \CII\ PDR layer. In the southeast,  where the C {\sc
ii} region expands into a dense, lumpy  molecular cloud ridge the expansion
is slowed down by the dense surrounding cloud, while it has expanded much
further to the northwest. There are places where the PDR shell has broken through
the cloud allowing radiation from the star to directly illuminate the cloud
further out.  
\label{fig_cartoon}}
\end{center}
\end{figure}

The CO(11--10) channel maps (Fig. \ref{fig_co11chanmap}), especially
the panel centered at 10.5 \kms, indicate the eastern boundary of hot gas
to be $\sim$ 90 -- 110\arcsec\ offset from the central star, HD\,37903.
This is seen more clearly in the CO(6--5) and CO(7--6) channel maps at
velocities 10.5 and 11 \kms\ (Fig.\,\ref{fig_co65chanmap} \&
\ref{fig_co76chanmap}), although the boundary is not completely smooth.
Figure \ref{fig_co11pvplots}  shows three position velocity plots over the
southeastern quadrant of the map, which cut through the molecular cloud
ridge in the east, south east and south.  These cuts  show that the CO
emission ends at $\sim$ 100\arcsec\ from the star, which is assumed  to be
the radius of the shell (0.19 pc). The cuts in CO(7--6) and CO(6--5)
(Fig.~\ref{fig_pvplots}) show the same. In these cuts, however, some faint
CO emission is seen all the way to the edge of the map, indicating that
there is some emission from the surrounding cloud even in CO transitions as
high as CO(7--6). At the boundary the CO(11--10) emission is centered at
10.5 \kms. The CO(11--10) lines are generally very narrow, $\sim$ 0.5 \kms,
suggesting that the shell is viewed tangentially.  The CO(7--6) and
CO(6--5) lines are broader ($>$1 \kms), probably because of contribution
from the more turbulent surrounding molecular cloud. The CO lines get
broader or double peaked closer to HD\,37903 (Fig.\,\ref{fig_co11pvplots}).
This is to be expected if the line of sight goes through the front and the
back of an expanding shell or a ridge inside the shell.  The narrow line
widths along the boundary of the CO(11--10) emission region suggest that
the emission resides in a thin shell. Figure~\ref{fig_cartoon} shows a
simplified picture of the expanding C {\sc ii} region surrounded by the hot
PDR region.

Not all the CO(11--10) is emitted from the  shell, however, because
there is definitely  structure in the CO emission even at a resolution of
23\arcsec. This emission appears to  originate in dense clumps, ridges and
filaments inside the \CII\ region or in the interface between the
surrounding dense, lumpy molecular cloud and the \CII\ region. At the
position of the CO peak (55\arcsec,$-$15\arcsec{}) the CO(11--10) emission
appears more like a northsouth ridge with a linear extent of 30\arcsec. The
CO lines are broad, 1.5 \kms,  and are not double peaked. This strong
CO(11--10) emission region, which also stands out in CO(7--6) and CO(6--5),
coincides with a prominent north south emission ridge, which is bright in
both PAH emission and vibrationally excited H$_2$.   The northsouth cut
goes through the eastern edge of the southern ridge (SR) \citep[see
e.g.][]{Sheffer11} at $\sim$ 75\arcsec\ south, which stands out as a clear
peak both in \CII\ and high-$J$ CO lines. The emission from the SR is
redshifted (V$_{lsr}$ $\sim$ 11 \kms{}) placing it near the backside of the
nebula.

The cloud in which the reflection nebula is embedded is much less dense
in the north and northwest, which is evident from the much fainter \thCO\
emission (Fig.\,\ref{fig_13co32chanmap}). Here the densities in the PDR
shell are too low to excite CO(11--10)\footnote{For gas at 100~K, the 
critical density for CO(11--10) is 4 $\times ~10^5$ cm$^{-3}$, while it is
$10^5$ and 7 $\times ~10^4$ cm$^{-3}$ for CO(7--6) and CO(6--5),
respectively \citep{Yang10}.}, but sufficiently high for CO(6--5) and
CO(7--6), which are both dominated by the hot gas in the PDR
(Fig.\,\ref{fig_intcoplots}). In the northwestern quadrant one can see
double split lines in both CO(7--6) and CO(6--5). Here the separation
between the two velocity features  is $\sim$ 2 \kms, suggesting that the
expansion velocity is $\sim$ 1 \kms. In the southeastern part of the nebula
the front and back side of the PDR is barely resolved in CO(11--10)  and
are separated in velocity by only $\sim$ 1 \kms, corresponding to an
expansion velocity into the dense ridge of $\sim$ 0.5 \kms.  These
expansion velocities are consistent with the linear extent of the C {\sc
ii} region, which is about twice as large in the  northwest  as it is to
the southeast.

\subsection{The origin of \CII\  and \crrl\ emission}

Although \CII\ is detected throughout the mapped region, the \CII\
emission is dominated  by the  PDR illuminated by HD\,37903
(Fig.\,\ref{fig_intplots}).  To the southeast, where the C {\sc ii} region 
expands into the dense lumpy molecular cloud ridge,  the \CII\ emission
extends out to $\sim$ 100\arcsec\ from HD\,37903. In the northwest, where
the surrounding molecular cloud  is more diffuse, the emission from the PDR
is fainter and extends out to $\sim$ 220\arcsec\ from the star.

\begin{figure}[h]
\begin{center}
\includegraphics[width=0.49\textwidth]{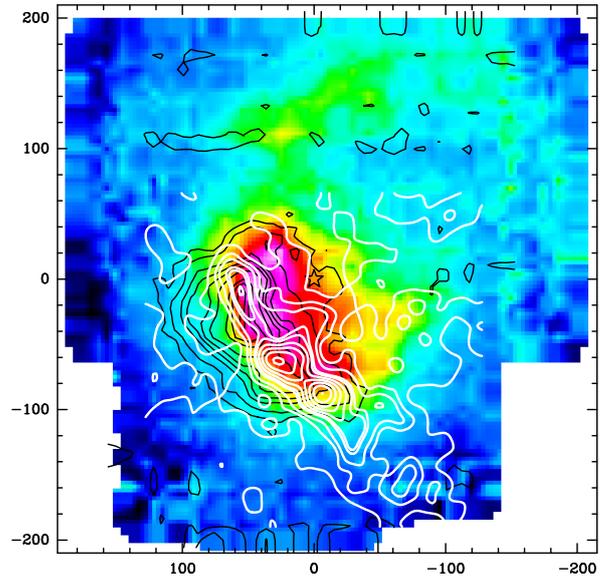}
\caption{Integrated intensity map of \CII\ (in color) overlayed with
CO(11-10) and C91$\alpha$ contours in black and white respectively.
For CO(11--10) the contours drawn are at 3, 5, 8, 10, 12, 15, 20, 22,
24, 26 times the value of $\sigma = 1.3$\,K\kms. For C91$\alpha$
contours are drawn at 5\%, 20\%, 30\%, 45\%, 50\% to 100\% in steps of
10\% of the peak intensity of 7.3~10$^{-3}$\,Jy/beam.
The star symbol marks the position of HD\,37903.
\label{fig_cpover}}
\end{center}
\end{figure}

The strongest  \CII\ emission is detected in the southeastern quadrant
where it is bounded by the dense molecular ridge (Fig. \ref{fig_cpover}). 
The position velocity cut through the southeastern ridge (Fig. 
\ref{fig_pvplots}) shows strong, somewhat blue-shifted emission inside the
molecular shell. The \CII\ lines, $\sim$ 2 -- 3 \kms\ wide, are  broader
than the CO(11--10) lines but roughly coincident in velocity.  Inside the
shell \CII\  also shows a fainter, broad,  blue-shifted wing extending to
velocities as low as 5 \kms.  The \CII\  channel maps (Fig.
\ref{fig_cpluschanmap}) show that this blue-shifted emission component is
present in most of the strong PDR emission region in the southeastern
quadrant. This blue-shifted emission can  easily be explained by a
photoevaporation flow from the dense PDR, suggesting that most of the PDR
emission is on the back side of the nebula.  There is also some fainter
($\lesssim$ 15 K) \CII\ emission to the west protruding out from the hot
molecular shell (Fig.\,\ref{fig_pvplots}). This emission, which is mostly
blue-shifted, probably originates in the warm surface layers of the
molecular cloud, where the densities are lower and the FUV radiation from
HD\,37903 can directly reach the cloud surface. Some of the \CII\ emission
outside the reflection nebula, $\sim$ 5 -- 10 K, may be  externally
illuminated by FUV radiation from the large H {\sc II}  region IC\,434,
which is located west of NGC\,2023.

\begin{figure*}[!]
\begin{center}
\includegraphics[width=\textwidth]{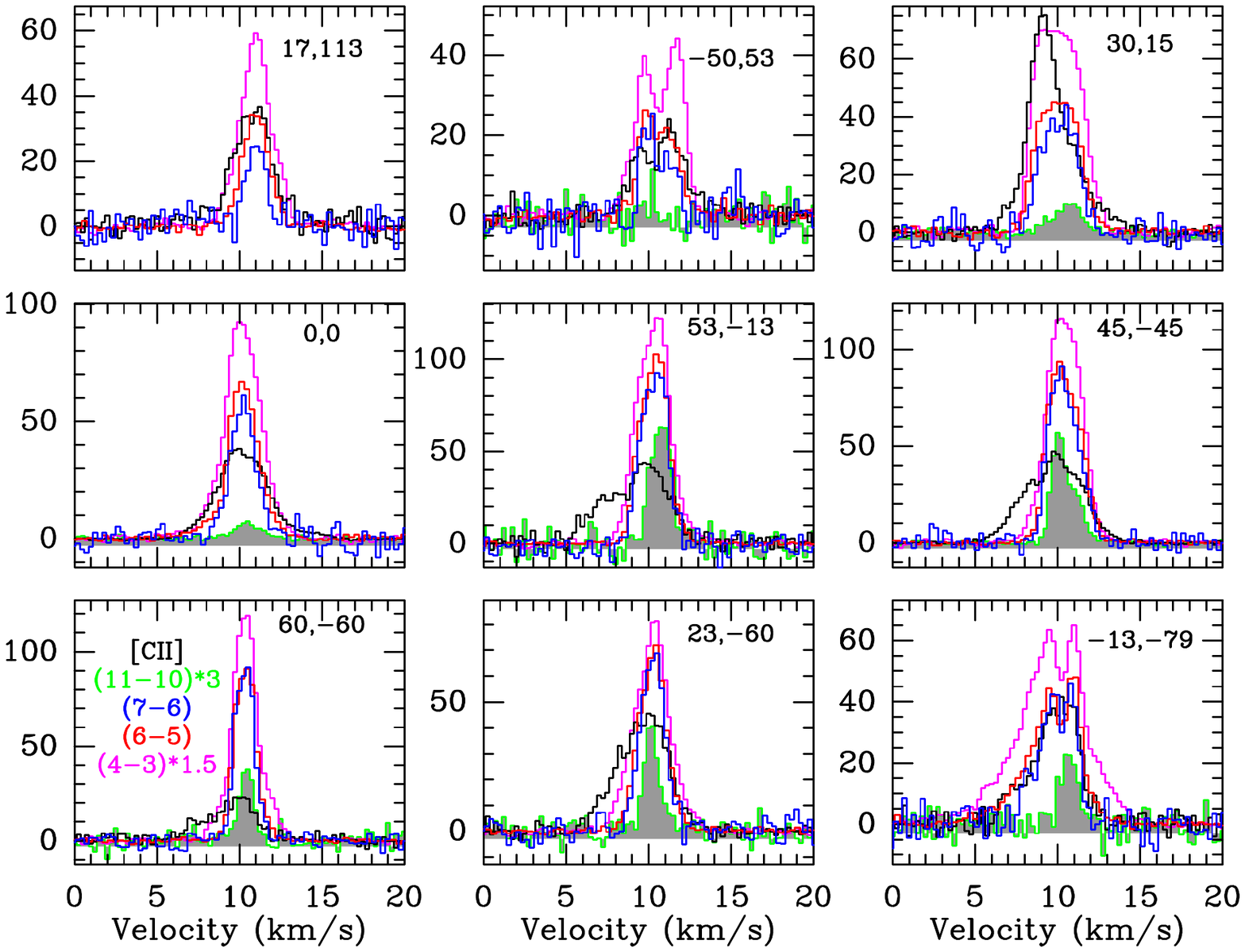}
\caption{Comparison of \CII, CO(11--10), CO(7--6), CO(6--5), and
CO(4--3) spectra at selected positions in the NGC\,2023 region. The
CO(4---3) spectra are scaled by a factor of 1.5  for clarity. CO(11--10) is
scaled by a factor of three and filled with greyscale. For clarity we
omitted CO(11--10) for the panel (17,113), because the spectrum is rather
noisy and a clear non-detection. 
\label{fig_selspec}}
\end{center}
\end{figure*}

In the northwest  the molecular cloud  has much lower density  than the
southeastern ridge. The northwestern position velocity plot
(Fig.\,\ref{fig_pvplots}) shows that  there are strong \CII{}-lines up to
$\sim$ 220\arcsec, which coincide with the PDR ridge. Inside this region
there are areas where the \CII\ and the high-$J$ CO lines are double peaked
suggesting that  they come from the front and back side of the expanding
\CII\ shell. The same double split lines are seen even in CO(3--2), see
Fig.\,\ref{fig_selspec}, at the offset $\sim$  -50\arcsec,+53\arcsec. The
\thCO\ line is dominated by emission from the surrounding cloud and the
column densities in the hot thin PDR shell are too low to be detected in
\thCO.  The velocity separation between the two velocity peaks (see
Table\,\ref{tab_n2023pos1} ), gives an expansion velocity of  $\sim$ 1
\kms, agreeing quite well with the linear extent of the C {\sc II} region
in the northwest, $\sim$ 220\arcsec, which is about twice of its extent to
the southeast. Faint \CII\ emission is also present outside the \CII\
shell. The \CII\ emission extends at least up to 230\arcsec\ from the
HD\,37903, and is red-shifted by several \kms. At the edge of our map the
velocity of the \CII\ emission is $\sim$ 12.4 \kms.

The \crrl\ emission mapped  by \citet{Wyrowski00} is only detected in
the southeast  (Fig.\,\ref{fig_cpover}), where the C {\sc ii} region
interacts with the dense molecular cloud. The \crrl\ emission is also
inside the hot molecular shell. The southernmost peak of the  \crrl\ 
ridge, however, is in a region where  no CO(11---10) emission is seen, nor
is there any evidence for a dense gas clump based on our  $^{13}$CO data.
Even though  \CII\ emission is seen, it is faint and there is no
enhancement in the emission at the position of the southern \crrl\ peak.
\subsection{Moderately optically thick \CII, evidence from \thCII{}-emission}

\begin{figure}[h]
\includegraphics[width=0.49\textwidth]{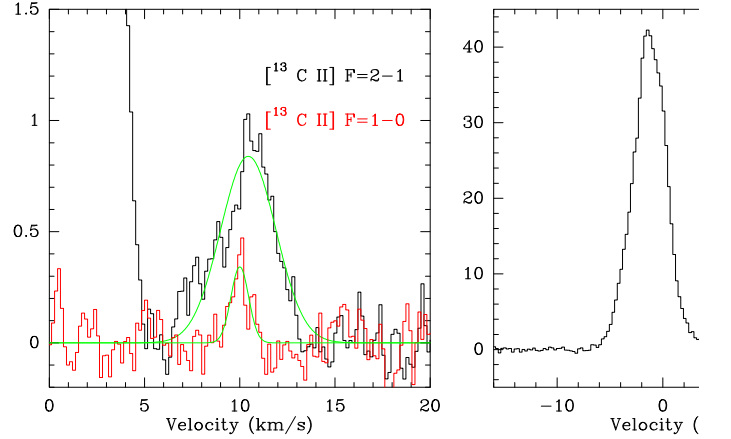}
\caption{\thCII\ spectra calibrated in main beam brightness temperature
and plotted on a common V$_{lsr}$ scale of \thCII. On the left is the
long integration HIFI spectrum toward HD\,37903. The F = 2--1 is  plotted
in black and F = 1--0 in red overlaid with Gaussian fits in green. To the
right is an average of all GREAT \CII\ spectra brighter than 35 K in the SE
quadrant of our map of NGC\, 2023. The F = 2-1 line is clearly visible at 
V$_{lsr}$ =   10.1 \kms, see text.
\label{fig_13cp}}
\end{figure}

If  \thCII\  is detected, the ratio of \thCII\ to \CII\ line intensity,
coupled with the $^{12}$C/$^{13}$C abundance  ratio, provides a measure of
the optical depth of the \CII\  emission \citep{Boreiko96}. Our GREAT
observations cover two of the three \thCII\ hyperfine lines: the F = 2--1
and F = 1--1, offset by +11.2 and +63.2 \kms\ relative to the \CII\
velocity, and  with relative intensities of 0.625 and 0.125 of the total
\thCII\ line intensity \citep{Cooksy86,Graf12,Ossenkopf13}. Although the
short integration times did not allow a clear detection of the \thCII\ F =
2--1 line in individual GREAT spectra, the F = 2--1 transition is
definitely present even in the average of all spectra in the map cube.
However, in a deep \CII\ spectrum\footnote{This is a 29 minute long load
switched observation from the {\it Herschel}  Open Time project 
OT1\_tvelusam\_1, retrieved from the {\it Herschel} science archive.}
observed toward HD\,37903 with HIFI on the {\it Herschel Space Observatory}
, the two strongest hyperfine lines of \thCII\ , the F = 2--1 at 1900.466
GHz, and F = 1--0 at 1900.950 GHz \citep{Ossenkopf13}, with relative
intensities of 0.625 and 0.250 respectively, are clearly visible
(Fig.\,\ref{fig_13cp}). The F = 1 --1 line is too faint to be detected in
any spectra. Using the fitted peak temperatures for the  \CII\  and 
\thCII\ hyperfine line transitions, the ratio of \CII\, and the sum of the
two \thCII\ lines, corrected for the relative intensity of the two
hyperfine transitions (0.875; Ossenkopf et al. 2013), is found to be 32.3.

To improve the signal to noise of the \thCII\ line in the GREAT data,
all positions in the SE quadrant where the  \CII\ intensity exceeds 35 K
were averaged. Figure \ref{fig_13cp} (right panel) shows this averaged
spectrum. In this  spectrum the \thCII\  F = 2 --1 transition has a peak
temperature of 1.28 K at a V$_{lsr}$ = 21.3 \kms, while the peak of the
\CII\ line is 41.7 K at 10.0 \kms. The difference in velocity, 11.3 \kms,
is in excellent agreement with the expected velocity separation of 11.2
\kms\ between the \CII\ line and the F = 2--1 transition of \thCII. The
width of the  \thCII\ line is $\sim$ 1.9 \kms, while the \CII\ line is
about twice as wide, 3.7 \kms. The ratio of the peak temperature between
\CII\, and \thCII, corrected for the relative intensity of the F = 2--1
line (0.625)  is $\sim$ 20.3. This ratio is very similar to what was
derived from the spectrum with HIFI towards HD\,37903 (see above). If the
\CII\ line is optically thin, the ratio is expected to be $\sim$ 70, the
isotope ratio of $^{12}$C/ $^{13}$C measured for Orion (see e.g.,
\citet{Ossenkopf13}). The \CII\ emission is therefore somewhat optically
thick with an optical depth of 1--2. All spectra in the northwestern
quadrant were also averaged.  Even here the \thCII\ line is present, but
only as a 3-$\sigma$ detection. However, the ratio between \CII\ and
\thCII\ is approximately the same. Analyzing other regions in the map give
similar results. The C {\sc II} region ionized by HD\,37903 is therefore
somewhat optically thick in \CII, since the  \thCII\ is readily detected in
all regions examined.

Since the  \CII\ emission is somewhat optically thick over the whole
nebula, this can be used to determine the excitation temperate for \CII.
Assuming a beam filling factor of one, and an optical depth of 2, the
observed brightness temperature at the position of HD\,37903, gives a lower
limit to T$_{ex}$ = 85 K. At the \CII\ peak, assuming the same optical
depth, one gets T$_{ex}$ = 130 K. Overall our observations suggest that
\CII\ has a similar excitation temperature as the CO in the PDR, perhaps
slightly warmer, or on the order of 90 -- 150 K. To improve this estimate a
few deep \CII\ integrations are needed  providing sufficient
signal-to-noise to enable a more  accurate determination of the optical
depth.

\subsection{HD\,37903}

HD\,37903 was found to have faint mid-IR excess emission from a
remnant disk or residual envelope by \citet{Mookerjea09}. A ``broad'' (1.9
\kms{}) but faint line in CO(11--10) is seen toward the star
(Table~\ref{tab_n2023pos1}). However, the position velocity cuts in
CO(11--10) (Fig.\,\ref{fig_co11pvplots}), which go through the star in
three different directions, show no enhancement at all at the position of
HD\,37903. Instead there is a red-shifted cloudlet at $\sim$ 11.5 \kms,
peaking $\sim$ 20\arcsec\ southeast  and south of the star. 
\citet{Wyrowski00} found weak extended  continuum  emission at  8 GHz (3.3
cm), with a peak southwest of HD\,37903 at ($-$10\arcsec, $-$12\arcsec{}),
which they interpreted as optically thin free-free radiation from gas
ionized by the B-star. This cloud could be associated with the ridge of
bright PAH emission which runs from northeast to southwest just south of
the star.  There is no clear \CII\  enhancement at this position, nor in
C(11--10) or in any of the low-$J$ CO maps, except that they all show
blue-shifted emission feature south west of the star, which peaks roughly
at ($-$10\arcsec,$-$18\arcsec{}) from the star. A possible scenario could
be that  a clump of gas is being pushed away by radiation pressure from the
strong FUV radiation close to the star, with the side facing the star
completely ionized while the emission further from the star is shielded and
still molecular.

\subsection{The physical conditions of the cloud surrounding NGC\,2023}

\begin{figure*}[t]
\begin{center}
\includegraphics[width=\textwidth]{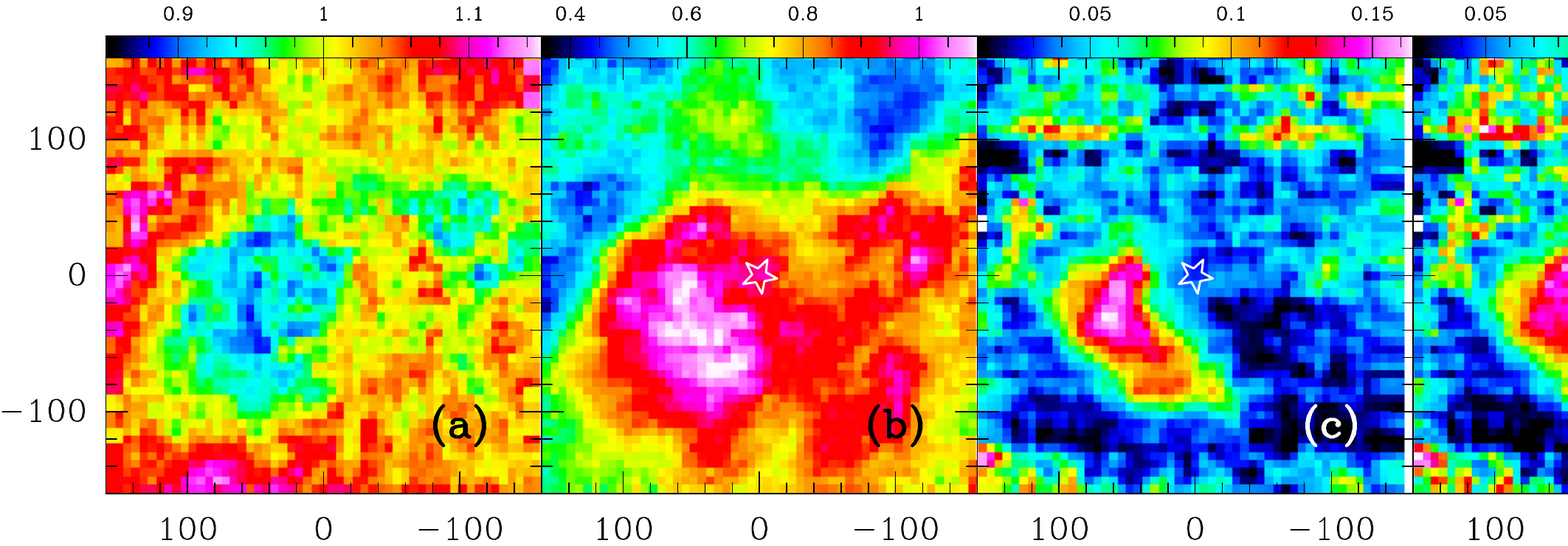}
\caption{Ratios of integrated intensity over the velocity range 9 -- 12
\kms.  From left to right we show: (a) CO(3--2)/CO(4--3), (b)
CO(6--5)/CO(4--3), (c) CO(11--10)/CO(6--5), and (d) CO(11--10)/CO(7--6).
The emission in CO(4--3) and CO(3--2) are dominated by the quiescent
colder molecular cloud, while the higher J transitions trace the hot
molecular shell.
\label{fig_linerats}}
\end{center}
\end{figure*}

The integrated intensity ratio CO(3--2)/CO(4--3) changes very little
between 0.9 and 1.1. This suggests that both the CO(4--3) and
CO(3--2) lines are optically thick at most positions in the cloud and
hence the brightness temperature is approximately equal to the gas
temperature. Using the peak brightness temperatures observed in CO(3--2)
we estimate the gas temperature of the quiescent cloud outside the C {\sc II} region
to be $\sim$ 40\,K. If the CO emission is clumped the gas temperatures could be 
somewhat higher. At the CO(3--2) emission peak there is a strong contribution 
from the hot PDR emission and the observed CO(3--2) brightness
temperature suggests a temperature of $\gtrsim$ 75~K.

In order to further constrain the kinetic temperature, density and
column density of the gas in the quiescent cloud and the hot molecular
shell a grid of models based on the non--LTE radiative transfer program
RADEX \citep{vdtak2007} have been compared with the observed line intensity
ratios.  Model inputs are molecular data from the LAMDA database
\citep{schoier2005} and CO collisional rate coefficients  from the work of
\citet{Yang10}.  RADEX predicts line intensities of a given molecule in a
chosen spectral range for a given set of parameters: kinetic temperature,
column density, H$_{2}$ density, background temperature and line width.
Unlike some of the lighter hydrides for which the excitation is dominated
by the far- and mid-infrared background radiation, the only radiation
affecting the excitation of CO molecules at (sub)-millimeter wavelengths is
the cosmic microwave background (CMB). Hence a value of 2.73\,K is assumed
as the background temperature for all calculations presented here. The
synthetic line ratios are calculated assuming line widths of 3\,\kms, which
is somewhat higher  than the observed line widths, but such line widths are
needed in order to capture the PDR emission throughout the nebula.

\begin{figure*}[!]
\begin{center}
\includegraphics[width=0.95\textwidth]{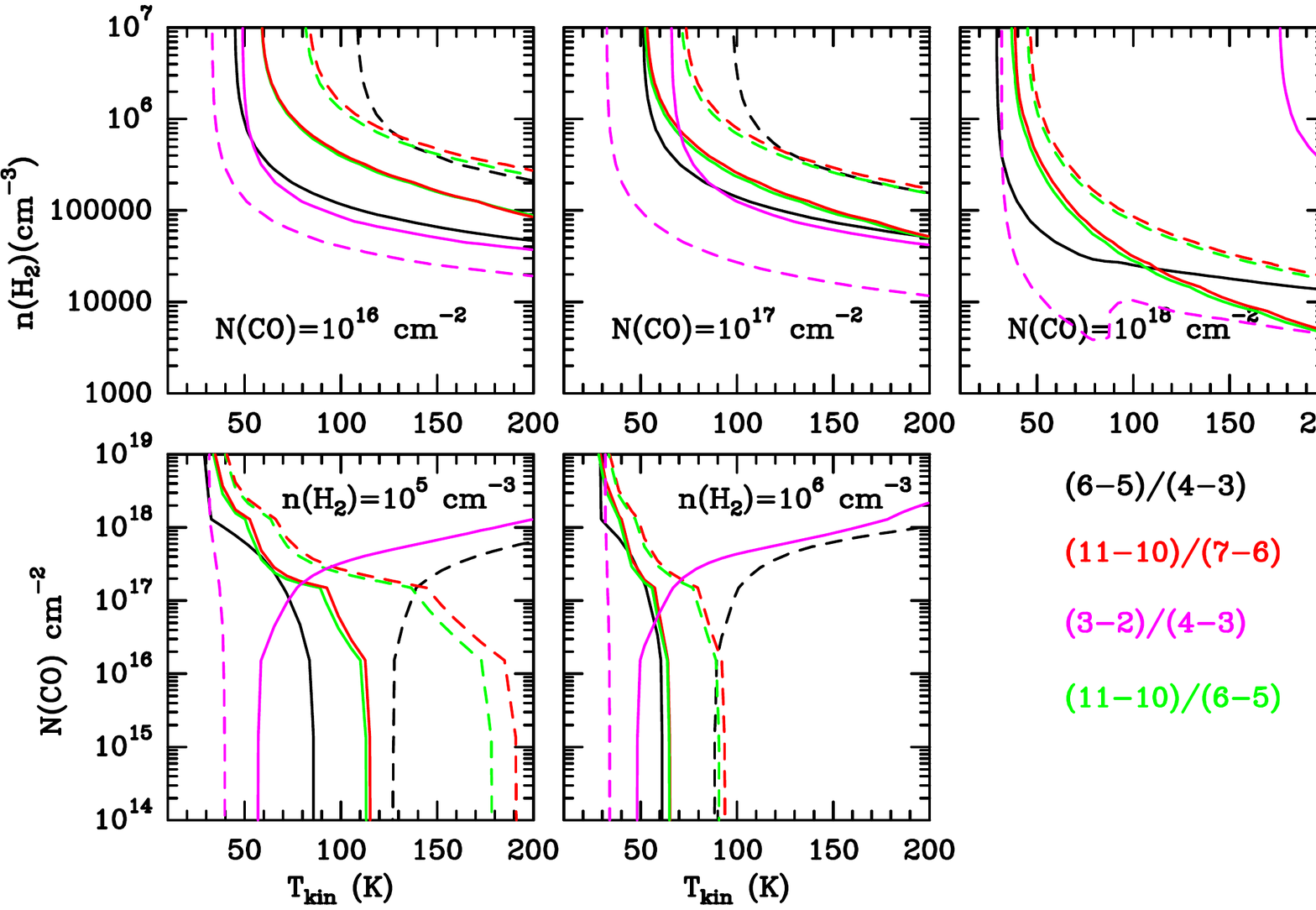}
\caption{Plot of intensity ratios based on synthetic models calculated
using RADEX. The contours correspond to intensity ratios indicated in
the panel as a function of the physical parameters. The ratios are color
coded, with the solid line representing the low limit and the dashed one
the high limit for each ratio as determined from the ratio plots in Fig.
\ref{fig_linerats}. The color coding and limits are as follows: black
(6--5)/(4--3) 0.9 (solid) -- 1.10 (dashed); red (11--10)/(7--6) 0.07
(solid) -- 0.20 (dashed):  green (11--10)/(6--5)  0.05 (solid) -- 0.16
(dashed);  magenta (3--2)/(4--3) 0.8 (solid) -- 1.1 (dashed).  The upper
row corresponds to models for which \Tkin\ and gas density ($n$(H$_2$)) are
varied keeping the column density ($N$(CO)) constant to the value
indicated in the panel.  In the lower row results models in which \Tkin\
and $N$(CO) are varied and $n$(H$_2$) are fixed to the values indicated in
the panel.
\label{fig_radex}}
\end{center}
\end{figure*}

Figure~\ref{fig_radex} shows  contours of the observed
CO(3--2)/CO(4--3), CO(6--5)/CO(4--3), CO(11--10)/CO(6-5) and
CO(11--10)/CO(7--6) ratios estimated as a function of {\it i}) density
($n$(H$_2$)) and kinetic temperature $T_{\rm kin}$, with $N$(CO) fixed) and
{\it ii}) $N$(CO) and $T_{\rm kin}$ as calculated by RADEX. The continuous
and dashed contours show the minimum and maximum values of the observed
ratios respectively.  While the intensity ratios do not constrain the
column densities at all, they uniquely demarcate the kinetic temperatures
and densities of the clouds which contribute to the emission of the CO
transitions being considered. As seen in Fig.\,\ref{fig_linerats} the
CO(3--2) \& CO(4--3) maps primarily trace the quiescent cloud.  Based on
the C$^{18}$O(2--1) observations by \citet{Jaffe90} and assuming that
C$^{18}$O is optically thin, the column densities in the surrounding
molecular cloud are as high as 5 $\times$ 10$^{18}$ -- 10$^{19}$ cm$^{-2}$.
 Therefore based on modeling with RADEX the \Tkin\ of the quiescent cloud
is estimated to be 35 -- 40\,K for densities of 10$^5$ to 10$^6$\,\cmcub.

The gas densities in the CO(11--10) emitting PDR shell have to be at
least 4 $\times$ 10$^5$ \cmcub, the critical density of CO(11--10). In the
north and northwestern part of the nebula, where CO(7--6) is readily
detected but not CO(11--10); the densities are somewhat lower, $\sim$
10$^5$ \cmcub. To get a rough idea of the CO column densities in the PDR
these densities are adopted and the width of the CO emitting layer is
assumed to be $\sim$ 1\arcsec. For a normal CO/H$_2$ abundance ratio of
10$^{-4}$  N(CO) is $\sim$ 10$^{17}$ cm$^{-2}$, which agrees well with what
\citet{Jaffe90} deduced from their observations. Since the ratios involving
CO(11--10) primarily trace the physical conditions of the hot molecular
shell, the \Tkin\ of  the shell  Is estimated to be $\sim$ 90 -- 120\,K for
gas densities of 10$^5$  -- 10$^6$\,\cmcub\ based on Fig.\,\ref{fig_radex}.

\subsection{Far-IR line cooling from \CII\ and CO}

It is well known that the fine structure lines of \CII\ 158 $\mu$m is
the dominant cooling line in low-density regions \citep{Hollenbach91}
providing $\sim$  0.3 \% of the total FIR (72 -- 196 $\mu$m) luminosity in
our Galaxy \citep{Stacey85}. Since our observations cover almost the whole
NGC\,2023 reflection nebula, it is possible to determine the \CII\
luminosity for a ``typical'' reflection nebula. Integrating the \CII\
emission from 4  -- 18 \kms\ over the C {\sc II} region, an area of 71100
arcsec$^2$, yields an integrated intensity of 8.86 $\times$ 10$^{-13}$
W~m$^{-2}$, corresponding to 3.4\Lsun\ for a distance of  350 pc. This
agrees well with the \CII\ luminosity estimated by \citet{Howe91}, who
estimated 3.8 \Lsun\ (corrected to a distance of 350 pc) from the area they
mapped with the KAO. NGC\,2023 was observed with PACS as part of the {\it
Herschel} Gould Belt survey \citep{Andre10}, but only at 70 and 160 $\mu$m.
However, the PACS wavelength coverage (60 -- 210 $\mu$m)
\citep{Poglitsch10} is fairly similar to FIR coverage used by
\citet{Stacey85}, allowing an estimate of the flux densities for the same
area used to determine the \CII\ luminosity. Since there is strong emission
from the surrounding molecular cloud, the integrated emission from the
reflection nebula was estimated by using a  background region to correct
for the emission from the surrounding cloud. From the PACS images the
background subtracted flux densities were found to be  3200 Jy and 3300 Jy
for 70 $\mu$m and 160 $\mu$m, respectively. Assuming a dust emissivity
index, $\beta$ = 1, the ratio of the 70 to 160 $\mu$m flux densities
correspond to a dust temperature of 33.3 K, which was used to estimate the
expected flux density at 100 $\mu$m, 4400 Jy. For such dust temperatures
the color correction of the PACS bands is negligible \citep{Muller11},
except for the 70 $\mu$m band, which requires a color correction of  $\sim$
5 \%. Therefore the effective bandwidths from the three PACS photometry
bands\footnote{Herschel PACS ICC Technical Note: The Bandwidth of the PACS
Photometric System, PICC-CR-TN-044, 2013} can be used to compute the FIR
luminosity for NGC2023, which comes to L$_{FIR}$ = 210 \Lsun. For NGC\.2023
the \CII\ luminosity is 1.6\% of the FIR luminosity, which sounds quite
plausible considering that \CII\ is much brighter in NGC\,2023 than in the
halos of molecular clouds.

Since  CO(11--10) was readily detected in the dense southeastern part
of the nebula, one can  make a rough estimate of how much cooling is
provided by CO. The  \CII, CO, and FIR luminosity are determined by
integrating all our images over the intense PDR emission in the
southeastern and central part of the nebula. This is essentially  the
region where CO(7--6) and CO(6--5) emission is very strong, (see
Fig.\,\ref{fig_co76chanmap} \& \ref{fig_co65chanmap}), and covers an area
of 31700 arcsec$^2$. For \CII\ a luminosity of 1.9 \Lsun\ is obtained,
i.e., almost 60\% of the \CII\ luminosity comes from the dense PDR. The
emission of CO lines is integrated over the velocity range  8.9 -- 12.6
\kms, which should cover all the emission from the hot PDR. For CO(11--10)
the luminosity  is 35.0 $\times$ 10$^{-3}$ \Lsun. For the lower $J$
transitions there is substantial contribution from the surrounding cloud.
Therefore the integrated emission was corrected using an adjacent reference
region. The derived luminosities for the 7--6, 6--5, 4--3, and 3--2
transitions are: 13.7, 10.8, 6.5 one 3.2 respectively; all in units of
10$^{-3}$  \Lsun. By estimating the  contribution of the CO transitions
which were not observed, results in a luminosity of $\sim$ 0.15 \Lsun\ for
CO transitions up to 11--10. The contribution from higher transitions is
more difficult to estimate, since there are no observations of any CO
transitions higher than 11--10. It is therefore assumed that higher $J$ CO
transitions contribute the same amount as all CO lines up to 11--10, which
is probably good to a factor of 2. The total CO luminosity  from the dense
PDR region is therefore $\sim$ 0.3 \Lsun, certainly not more than 0.5
\Lsun. Even though cooling from CO is by no means negligible, it is still a
factor of four lower than the cooling from \CII. Background subtracted flux
densities of the PACS 70 and 160 $\mu$m images for the southeastern region
are 2650 and 2400 Jy  at 70 and 160 $\mu$m , respectively. In this region
the dust temperature is 37 K resulting in a flux density of 3400 Jy at 100
$\mu$m. The color corrections are now negligible and the FIR luminosity is
160 \Lsun, i.e., the cooling efficiency from \CII\  is lower in very dense
regions, or only about 1.1\% of the FIR luminosity.

These results can be compared to the Orion Bar, another well-studied
bright PDR illuminated by a high mass star and often considered a
prototypical PDR template for high mass star forming regions.
\citet{BernardSalas12} find that in the dense Orion Bar PDR the \OI\ 63
$\mu$m lined accounts for 72\% of the total line cooling, with \CII\
contributing less than 18\% and CO lines $\sim$ 5\%, i.e., the contribution
from CO is $\sim$ 25\% of the \CII\ luminosity. The Orion Bar is therefore
similar to the dense PDR emission seen in the southeastern part of
NGC\,2023 in the sense that the cooling from CO is still lower than from
\CII, but not by a huge amount. In contrast cooling from molecular lines
completely dominate over atomic fine structure lines lines in high-mass
 protostars, where cooling from \CII\ is completely negligible
\citep{Karska14}.

\begin{table*}
\caption{Gaussian fits to spectra at selected positions. The CO(7--6),
CO(6--5) and CO(4--3) spectra have been convolved to the same spectral
resolution as \CII. CO(11--10) upper limits are 3$\sigma$ for a 1 \kms\
wide line. All offsets are measured relative to the position of
HD\,37903. 
\label{tab_n2023pos1}}
\centering
{\scriptsize
\begin{tabular}{llrrrrrrrrr}
\hline\hline
Offset & Tracer & $\int T_{mb} dv$\phantom{0} & $V_{\rm LSR}$\phantom{00} & $\Delta V$\phantom{00}
&  $\int T_{mb} dv$\phantom{0} & $V_{\rm LSR}$\phantom{00} & $\Delta V$\phantom{00}
&  $\int T_{mb} dv$\phantom{0} & $V_{\rm LSR}$\phantom{00} & $\Delta V$\phantom{00}\\
(\arcsec,\arcsec) && (K\,\kms) & (\kms{}) & (\kms) & (K\,\kms) & (\kms) & (\kms)
& (K\,\kms) & (\kms) & (\kms)\\
\hline
& & &\\
(0.0,0.0) & $^{13}$CO(3--2)   &  31.6$\pm$0.3 & 10.25$\pm$0.01 &  1.59$\pm$0.02 &&&\\
         &   CO(3--2)        &  155.4$\pm$2.3 & 10.23$\pm$0.02 &  2.52$\pm$0.05 &&&\\
         &   CO(4--3)        &  165.8$\pm$0.6 & 10.22$\pm$0.01 &  2.55$\pm$0.01 &&&\\  
         &   CO(6--5)        &  131.3$\pm$0.5 &  10.25$\pm$0.01 &  2.23$\pm$0.01 &&& \\
         &   CO(7--6)        &   84.7$\pm$0.9 &  10.24$\pm$0.01 &  2.01$\pm$0.03 &&& \\
         &   CO(11--10)      & 4.3$\pm$0.2 & 10.51$\pm$0.05 & 1.89$\pm$0.13 &&&\\
         &   \CII{}$^a$           & 130.3$\pm$1.3 & 10.36$\pm$0.02 & 3.45$\pm$0.05&  6.3$\pm$1.01 & 9.52$\pm$0.04 &  1.00$\pm$0.09\\
\hline
(30,15)  & $^{13}$CO(3--2)   &  8.1$\pm$0.4 & 9.09$\pm$0.03 &  1.06$\pm$0.05 & 23.9$\pm$0.1 & 10.45$\pm$0.01 & 1.74$\pm$0.01\\
         &   CO(3--2)        &  47.7$\pm$0.5 & 9.42$\pm$0.01 &  1.30$\pm$0.01 &  103.7$\pm$0.7  &  10.77$\pm$0.01 &  2.32$\pm$0.02  \\
         &   CO(4--3)        &  49.1$\pm$0.0 & 9.45$\pm$0.01 &  1.26$\pm$0.01 & 112.0$\pm$0.5  &  10.78$\pm$0.01 &  2.23$\pm$0.02 \\
         &   CO(6--5)        &  36.6$\pm$4.2 &  9.49$\pm$0.25 &  1.11$\pm$0.25 & 96.2$\pm$4.2 &  10.66$\pm$0.25 &  2.00$\pm$0.25\\
         &   CO(7--6)        &  21.8$\pm$4.1 &  9.62$\pm$0.03 &  1.07$\pm$0.09 & 65.6$\pm$4.2  & 10.56$\pm$0.06 &  2.04$\pm$0.05 \\
         &   CO(11--10)      & 0.91$\pm$0.4 &  9.21$\pm$0.10 & 0.71$\pm$0.29 &  6.0$\pm$0.4 & 10.75$\pm$0.06 & 1.66$\pm$0.15\\
         &   \CII{}$^b$           & 159.7$\pm$3.9& 9.10$\pm$0.02 &  2.02$\pm$0.04 & 51.5$\pm$4.1 & 11.19$\pm$0.06 & 1.97$\pm$0.12\\
\hline
(45,-45) & $^{13}$CO(3--2)   &  56.0$\pm$0.4 & 10.09$\pm$0.01 &  1.34$\pm$0.01 & & & \\
         &   CO(3--2)        &  186.6$\pm$2.7 & 10.39$\pm$0.02 &  2.44$\pm$0.04 &&&\\
         &   CO(4--3)        &  205.9$\pm$3.2 & 10.41$\pm$0.02 &  2.41$\pm$0.04 &&&\\
         &   CO(6--5)        &  85.8$\pm$3.5 &  9.85$\pm$0.02 &  1.36$\pm$0.02 & 92.3$\pm$3.5  &  11.00$\pm$0.03 &  1.55$\pm$0.03\\
         &   CO(7--6)        &  19.4$\pm$0.5 &  9.70$\pm$0.02 &  0.97$\pm$0.03 & 112.9$\pm$0.5  & 10.54$\pm$0.01 &  1.93$\pm$0.02\\
         &   CO(11--10)      & 10.9$\pm$1.3 &  9.95$\pm$0.13 & 0.78$\pm$0.13 & 19.1$\pm$1.3 & 10.69$\pm$0.13 & 1.76$\pm$0.13\\
         &   \CII\           & 55.0$\pm$2.1 & 8.33$\pm$0.09 &  3.41$\pm$0.09 & 125.2$\pm$2.1 & 10.42$\pm$0.09 & 3.27$\pm$0.09\\
\hline
(60,-60) & $^{13}$CO(3--2)   &  42.9$\pm$2.2 & 10.18$\pm$0.25 &  0.98$\pm$0.25 & 6.7$\pm$2.2 & 11.17$\pm$0.25 & 1.34$\pm$0.25\\
         &   CO(3--2)        &  102.4$\pm$4.9 & 10.38$\pm$0.25 &  1.61$\pm$0.25 & 49.1$\pm$4.9 & 10.60$\pm$0.25 & 3.57$\pm$0.25 \\
         &   CO(4--3)        &  78.2$\pm$2.6 & 10.34$\pm$0.01 &  1.36$\pm$0.02 & 83.0$\pm$2.5 & 10.49$\pm$0.01 & 2.79$\pm$0.05\\
         &   CO(6--5)        &&&&  133.0$\pm$0.3 &  10.37$\pm$0.01 &  1.58$\pm$0.01\\
         &   CO(7--6)        & &&& 96.1$\pm$0.1 &  10.35$\pm$0.01 &  1.41$\pm$0.01\\
         &   CO(11--10)     &&& & 15.7$\pm$0.9 &  10.43$\pm$0.03 & 1.03$\pm$0.07\\
         &   \CII\           & 21.0$\pm$6.6 & 7.51$\pm$0.27 &  2.01$\pm$0.57 &  56.4$\pm$6.5 & 10.02$\pm$0.12 & 2.26$\pm$0.26\\
\hline
  (-13,-79) & $^{13}$CO(3--2)   &  51.2$\pm$0.5 & 10.03$\pm$0.01 &  2.01$\pm$0.02 & &&\\
         &   CO(3--2)        &  137.2$\pm$0.1  &  10.07$\pm$0.00 &  4.09$\pm$0.00  &  71.4$\pm$0.1 & 8.51$\pm$0.00 &  7.44$\pm$0.00\\
         &   CO(4--3)        &  152.2$\pm$0.1 & 10.06$\pm$0.00 &  4.09$\pm$0.00 & 50.5$\pm$0.3 &  7.75$\pm$0.01 &  6.65$\pm$0.10 \\
         &   CO(6--5)        &  50.9$\pm$3.5  &  8.49$\pm$0.25 &  3.72$\pm$0.25 &  31.1$\pm$3.5 &  9.42$\pm$0.25 &  1.27$\pm$0.25 &48.9$\pm$3.5 &
	 10.97$\pm$0.25 & 1.47$\pm$0.25\\
         &   CO(7--6)        &  24.6$\pm$2.1 &  8.19$\pm$0.12 &  2.88$\pm$0.29 & 31.9$\pm$0.4 & 9.55$\pm$0.03 &  1.37$\pm$0.04 &
	 36.0$\pm$1.4 & 10.85$\pm$0.02 & 1.30$\pm$0.00$^c$ \\
         &   CO(11--10)     &&& &  &&&9.8$\pm$1.0 & 10.71$\pm$0.06 & 1.30$\pm$0.15\\
         &   \CII\    &&&       & 64.2$\pm$1.9 & 8.85$\pm$0.06 &  3.32$\pm$0.22 & 86.0$\pm$1.8 & 10.61$\pm$0.04 &  2.36$\pm$0.09\\

\hline
(53,-13) & $^{13}$CO(3--2)   &  47.5$\pm$0.5 & 10.03$\pm$0.01 &  1.64$\pm$0.02 &&&\\
         &   CO(3--2)        &  46.7$\pm$0.7 & 10.00$\pm$0.00$^c$ &  2.41$\pm$0.06 & 139.8$\pm$0.1 & 10.40$\pm$0.01 & 2.46$\pm$0.02 \\
         &   CO(4--3)        &  66.4$\pm$0.7 & 10.00$\pm$0.00$^c$ &  2.40$\pm$0.04 & 135.7$\pm$0.3 & 10.45$\pm$0.01 & 2.26$\pm$0.02 \\
         &   CO(6--5)        &  37.0$\pm$1.1 &  10.00$\pm$0.00$^c$ &  2.17$\pm$0.03 & 140.9$\pm$1.1 & 10.43$\pm$0.01 & 1.988$\pm$0.01\\
         &   CO(7--6)        &&&& 129.0$\pm$0.8 &  10.37$\pm$0.01 &  2.00$\pm$0.01 \\
         &   CO(11--10)      &&&& 34.7$\pm$1.8 & 10.77$\pm$0.04 & 1.46$\pm$0.09 \\
         &   \CII\           & 50.6$\pm$7.8 &  7.13$\pm$0.15 &  2.30$\pm$0.28 & 126.9$\pm$8.2 & 10.03$\pm$0.09 & 2.83$\pm$0.18\\  
\hline
(23,-60) & $^{13}$CO(3--2)   &  46.8$\pm$0.4 &  9.99$\pm$0.01 &  1.31$\pm$0.02 &&&\\
         &   CO(3--2)        &  118.4$\pm$0.7 & 10.00$\pm$0.00$^c$ &  2.94$\pm$0.02 & 17.1$\pm$0.5 & 10.35$\pm$0.01 & 1.10$\pm$0.00$^c$ \\
         &   CO(4--3)        &  115.2$\pm$0.7 & 10.00$\pm$0.00$^c$ &  2.82$\pm$0.02 & 21.8$\pm$0.5 & 10.54$\pm$0.13 & 1.10$\pm$0.00$^c$ \\
         &   CO(6--5)        &  115.8$\pm$0.0 &  10.23$\pm$0.01 &  1.92$\pm$0.01 & &&\\
         &   CO(7--6)        &  90.3$\pm$0.8 &  10.21$\pm$0.01 &  1.75$\pm$0.02 &&&\\
         &   CO(11--10)      & 15.0$\pm$1.2 & 10.16$\pm$0.04 & 1.10$\pm$0.10 &&&\\
         &   \CII\           & 112.4$\pm$2.7 & 10.27$\pm$0.03 & 2.49$\pm$0.09  & 50.9$\pm$1.8 &  8.19$\pm$0.05 &  2.42$\pm$0.17\\
\hline
(-50,53) & $^{13}$CO(3--2)   &  26.7$\pm$1.0 & 10.30$\pm$0.04 &  2.03$\pm$0.10 & &&\\
         &   CO(3--2)        &  35.1$\pm$4.8 & 10.30$\pm$0.00 &  3.20$\pm$0.00 &  18.1$\pm$4.8  &  9.64$\pm$0.25 &  1.30$\pm$0.00 & 37.1$\pm$4.8 &
	 11.72$\pm$0.25 & 1.40$\pm$0.25\\
         &   CO(4--3)        &  19.3$\pm$1.7 & 10.30$\pm$0.00 &  3.28$\pm$0.38 & 30.9$\pm$1.4 &  9.78$\pm$0.02 &  1.52$\pm$0.07 & 37.0$\pm$1.3 &
	 11.71$\pm$0.03 & 1.35$\pm$0.05\\
         &   CO(6--5)        &  10.2$\pm$1.1 & 10.30$\pm$0.00 &  3.00$\pm$0.00 & 24.4$\pm$0.7  &  9.96$\pm$0.02 &  1.24$\pm$0.01 & 21.5$\pm$0.1 &
	 11.47$\pm$0.02 & 1.30$\pm$0.00$^c$\\
         &   CO(7--6)       &&& &  17.5$\pm$1.5 &  9.95$\pm$0.04 &  1.18$\pm$0.10 & 13.7$\pm$1.2  & 11.22$\pm$0.07 &  1.30$\pm$0.00$^c$ \\
         &   CO(11--10)      &   $<$1.9      &  \ldots        &  \ldots  &&&&&&\\
         &   \CII\           & &&&18.6$\pm$4.3 & 9.24$\pm$0.11 &  1.35$\pm$0.25 & 52.7$\pm$5.2 & 11.39$\pm$0.09 & 2.21$\pm$0.27\\ 
\hline
(17,113) & $^{13}$CO(3--2)   &  30.5$\pm$0.4 & 11.10$\pm$0.01 &  1.81$\pm$0.03 &&&\\
         &   CO(3--2)        &  91.2$\pm$0.5 & 11.00$\pm$0.01 &  2.33$\pm$0.01 &&&\\
         &   CO(4--3)        &  92.0$\pm$0.6 & 10.99$\pm$0.01 &  2.30$\pm$0.02 &&&\\
         &   CO(6--5)        &  59.7$\pm$0.3 &  10.98$\pm$0.01 &  1.92$\pm$0.01 &&&\\
         &   CO(7--6)        &  28.3$\pm$0.9 &  10.94$\pm$0.02 &  1.55$\pm$0.06 &&&\\
         &   CO(11--10)      & $<$2.9 & \ldots& \ldots &&&\\
         &   \CII\           & 101.2$\pm$3.4 & 10.82$\pm$0.04 &  2.72$\pm$0.11 &&&\\
\hline
\hline
\end{tabular}
{\noindent $^a$ Red-shifted wing (3.5 K \kms{}) ignored in Gaussian fit}\\
{\noindent $^b$ Both blue- and red-shifted wing ( 11.8 K \kms{}) ignored in Gaussian fit}\\
{\noindent $^c$ Not fitted, i.e., kept constant}
}
\end{table*}

\subsection{PDR modeling}

\subsubsection{Choice of positions in the PDR}

The close agreement in morphology between the integrated \CII\ emission
and fluorescent H$_2$ and PAH emission (Fig.~\ref{fig_intplots},
Section~\ref{sect:morphology}) confirms that dominant part of the \CII\
emission arises from the PDR illuminated by HD\,37903. This agrees well
with the PDR modeling by \citet{Kaufman06}, who showed that at least 85\%
of the \CII\ emission originates in the PDR. We find that the high-$J$ CO
lines, particularly CO(11--10) and CO(7--6), are also dominated by emission
from the PDR (Section~\ref{sect:coshell}), although there is still some
contribution ($\sim$10 \Kkms{}) from the surrounding cloud in CO(7--6)  in
the dense southeastern ridge. It is difficult to estimate the contribution
from the surrounding cloud, especially since the densest part of the
surrounding molecular cloud coincides with the PDR, as seen for example
from the C$^{18}$O(2--1) map by  \citet{Jaffe90}. By integrating over the
CO(7--6) emission from 9 -- 12 \kms\ inside the southern ridge and
estimating the background from a comparison region outside the nebula, it
is found that about half of the CO(7--6) emission and slightly more of the 
CO(6--5) comes from the surrounding molecular cloud. In the northwestern
part of the nebula the cloud emission is negligible in both CO(6--5) and
CO(7--6). Therefore \CII\  and CO(11--10) can be used to explore the
physical conditions in the southwestern quadrant, while results from
CO(7--6) and CO(6--5) have to be used with caution. For the rest of the
nebula these transitions work fine, while the lower-$J$ CO transitions are
dominated by the surrounding cloud. Since the physical conditions vary
quite of lot, nine positions have been selected for modeling,  located in
different parts of the reflection nebula.  These include: (0\arcsec,
0\arcsec{}),  the position of HD\,37903; (30\arcsec, 15\arcsec{}), the
\CII\ emission peak; (53\arcsec, $-$13\arcsec{}), the CO(11--10) as well as
C91$\alpha$ peak; (17\arcsec, 113\arcsec{}), the northern \CII\ peak;
(45\arcsec, $-$45\arcsec{}), a position in the CO(11--10) ridge;
($-$49\arcsec, 53\arcsec{}), a relatively benign position in the
northwestern quadrant, showing PDR emission from both the front and back
side of the PDR shell; (23\arcsec, $-$60\arcsec), centered on the
C91$\alpha$ clump no. 2, ($-$13\arcsec, $-$79\arcsec), the SR position,
which is the most extensively modeled PDR position in NGC\,2023, and
(60\arcsec, $-$60\arcsec), peak of the low-$J$ CO emission.  The selected
positions are discussed in more detail in  Appendix~\ref{sect:selpos}. The
spectral line profiles of all the selected positions are shown in
Fig\,\ref{fig_selspec}, and the line parameters derived from Gaussian fits
(described in the following text) are given in Table~\ref{tab_n2023pos1}.

\subsubsection{Determination of line parameters}

Since the spatial resolution of observed images varies from 24\arcsec\
for CO(11--10) down to 8\farcs2 for CO(7--6),  the maps have been smoothed
by re-gridding all maps in CLASS to the same angular resolution and on a
common grid. In order not to miss out on the considerable structure that
has been detected in our maps on all angular scales, all maps were smoothed
to 16\farcs1, the resolution of the \CII\ map, and CO(11--10) was left at
its original resolution (24\farcs2), since CO(11--10) is detected only in
the southeastern quadrant of the C {\sc II} region.  CO(3--2) and
\thCO{}(3--2) were also left at their native angular resolution, since
neither of them are good  tracers of PDR emission  and they are not used in
the PDR analysis.  Smoothing introduces relatively minor additional
calibration uncertainty, since all spectra were calibrated in T$_{mb}$
based on observations of Jupiter, which is an extended source with a size
of $\sim$ 45\arcsec\ and hence encompasses most of the power of the error
beam.

As noted earlier the line emission from the reflection nebula NGC\,2023
show evidence of emission from the quiescent cloud, from the PDR,  from the
high velocity molecular outflows and also from wings due to photo
evaporation flows (seen in \CII\ spectra). For the analysis of the PDR
emission most of the emission from the quiescent cloud has been excluded by
considering only CO(6--5), CO(7--6) and CO(11--10) along with the relevant
component of the \CII\ emission.  Molecular outflows are also excluded
except for the southern ridge (SR) position at ($-$13\arcsec,$-$79\arcsec).
The  extraction of the emission primarily due to the PDR component was
achieved using multi-component Gaussian fits to all the observed molecular
lines and \CII\ for all selected positions.

A stepwise approach was followed to disentangle the different velocity
components contributing to the observed line emission. Since \thCO{}(3--2)
is a good tracer of the quiescent surrounding molecular cloud
single-component Gaussian fits to the \thCO{}(3--2) line was performed to
obtain the velocity of the quiescent cloud. Similarly CO(11--10), if
present, and CO(7--6) spectra were used to identify the velocity of the hot
CO in the PDR by fitting a one- or two- (depending on whether the two
components arising from the front and the back of the PDR are discernible)
velocity component Gaussians to these spectra. The two components often
blend in the southeastern quadrant. The CO(6--5) spectra were fitted
assuming contributions from both the quiescent cloud and the PDR, and in
most cases the best results were obtained by locking the velocity of the
quiescent cloud to the value determined from \thCO{}(3--2).  The lower $J$
CO lines, CO(3--2) and CO(4--3) are trickier, because especially in the
southeastern quadrant they suffer from strong self-absorption by the
surrounding cloud. Reasonable fits to the low-$J$ CO lines typically
required two or three components by locking the quiescent cloud component. 
However, due to overlap in the velocity of the emission from the quiescent
cloud and the PDR, even when a good fit has been obtained for the low-$J$
CO lines, the intensities may not accurately represent either the cloud
emission or the PDR emission. Thus both CO(3--2) and CO(4--3) were excluded
in our comparison of the observations with the PDR models. Finally the
\CII\ line profiles were fitted using the high-$J$ CO emission as a
guideline, although it is noted that the \CII\ PDR emission is always
offset by  $\sim$ 0.2 -- 0.5  \kms{} from the hot CO emission and the
lines are generally broader. Additionally \CII\ often show high velocity
non-gaussian wings from photo-evaporations flows  which have no
molecular counterpart. These were masked out out during the fitting and
their contribution was determined by taking the difference between the
total line integral and the sum of the area covered by the Gaussian fits. 
The results of the Gaussian fits are given in Table ~\ref{tab_n2023pos1}.

\subsubsection{Comparison with plane parallel PDR models} 
\label{sect:PDRmodeling}

In order to estimate the physical conditions in the photon dominated
regions (PDRs) contributing to the emission at the selected positions the
observed line intensity ratios are compared with the results of the model
for PDRs by \citet{Kaufman06}. The physical structure in these models is
represented by a semi-infinite slab of constant density, which is
illuminated by FUV photons from one side. The model takes into account the
major heating and cooling processes and incorporates a detailed chemical
network. Comparing the observed intensities with the steady-state solutions
of the model, allows for the determination of the gas density of H nuclei,
$n_{\rm H}$, and  the FUV flux (6~eV $\leq$ h$\nu$ $<$ 13.6~eV), $G_0$,
measured in units of the Habing (1968) value for the average solar
neighborhood FUV flux, 1.6$\times 10^{-3}$ ergs\,cm$^{-2}$\,s$^{-1}$. As
noted already in this paper the relative contribution of the cloud and the
PDR to the total emission depends on the observed tracer and the location
in the nebula. Since the primary focus is to characterize the PDR in
NGC\,2023, only high-$J$ CO lines like (6--5), (7--6), (11--10) and \CII\
line intensities corresponding to the velocity components identified to be
associated with the PDR are being used. In Table~\ref{tab_pdrout} the
component number refers to the first, second or third velocity component of
Table~\ref{tab_n2023pos1}. Appendix~\ref{sect:selpos} discusses in more
detail the positions and velocity components which were modeled.

\begin{figure*}[!]
\begin{center}
\includegraphics[width=\textwidth]{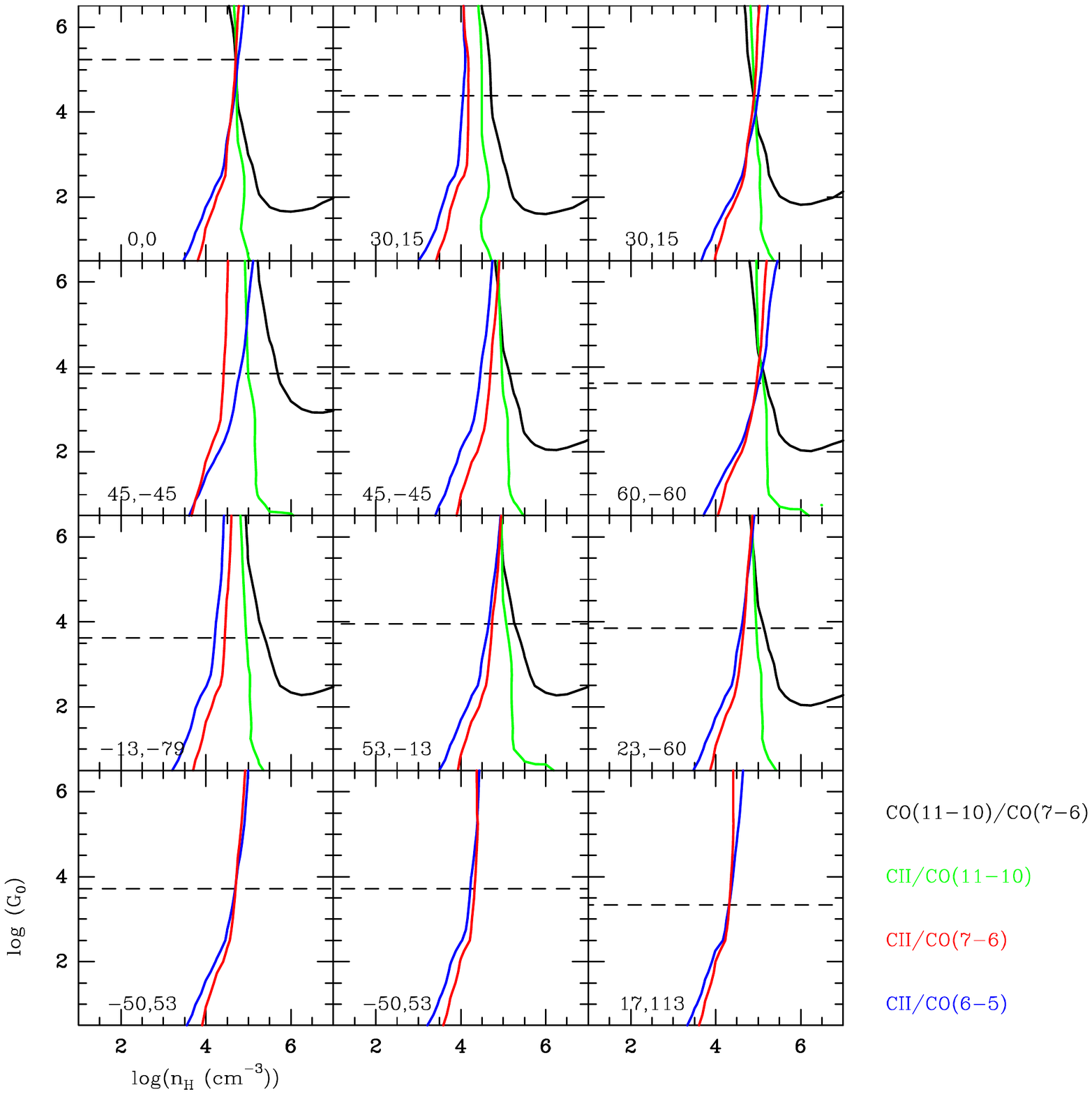}
\caption{Contours of observed intensity ratios plotted on the
intensity predictions as a function of hydrogen density $n_{\rm H}$ and
FUV radiation field ($G_0$) from PDR models, for the selected positions 
in NGC\,2023. The horizontal dashed lines correspond to FUV estimated 
using the stellar parameters as described in the text.
\label{fig_pdrmod}}
\end{center}
\end{figure*}

\begin{table}[h]
\caption{Results of comparison of observed intensity ratios with PDR model
calculations. Positions are given as offsets in arcseconds relative to HD\,37903.
\label{tab_pdrout}}
\begin{tabular}{ccccccc}
\hline
\hline
Position & Component \# & $n_{\rm H}$ & $G_0^a$\\
& & 10$^4$ \cmcub & 10$^3$(Habing Field)\\
\hline
(0,0) & 1 & 5.2 & 175\\
(30,15) & 1 & 1.2--5.2 & 24.7\\
(30,15) &  2 & 8.8 & 24.7\\
(45,-45) & 1 & 2.7--49.8 & 7.2\\
(45,-45) &  2 & 3.0--14.2 & 7.2\\
(60,-60) & 2 & 11.7 & 4.1\\
(-13,-79) & 3 & 1.7--23.1 & 4.2\\
(53,-13) & 2 & 4.5--19.0 & 9.2\\
(23,-60) & 1 & 3.9--13.6 & 7.2\\
(-50,53) &  2 & 5.2 & 5.3     \\
(-50,53) &  3 & 2.0 & 5.3\\
(17,113) & 1 & 2.1 & 2.2\\
\hline
\hline
\end{tabular}
$^a$ Estimated from stellar parameters as explained in the text.
\end{table}

Figure\,\ref{fig_pdrmod} shows results of comparison of the observed
intensity ratios \CII/CO(6--5), \CII/CO(7--6), \CII/CO(11--10) and
CO(11--10)/CO(7--6) with the predictions of the PDR models. It is assumed
that the beam-filling factors of the PDR components of the different
tracers used are identical and hence intensity ratios do not depend on the 
beam filling.  Each panel in Fig.\,\ref{fig_pdrmod} shows the observed
intensity ratios of different tracers as contours in different colors. The
strength of the FUV radiation in the region has been derived using the same
method as described by \citet{Sheffer11}. For this purpose a total FUV
luminosity of 4.0$\times 10^3$\,\lsun\ \citep{Parravano03} is used, which
is appropriate for a B2\,V star of 9\,\msun\  \citep{Trundle07,Hohle10} and
 the distance to NGC\,2023 to taken to be D=350\,pc. In each panel the
estimated FUV field strength is marked by dashed horizontal lines. Since
projected distances to the various positions in the cloud are being used,
the estimated FUV field will be an upper limit and varies between
2.2$\times 10^3$ and 1.7$\times 10^5$ in units of Habing field. There is no
way to know how close the PDR interface is at  the position of HD\,37903,
but it is probably no closer than 0.025 pc (i.e. 15\arcsec\ radius). 
Overall, all the ratios constrain the densities to values between 10$^4$
and 10$^{4.5}$\,\cmcub\ corresponding to the range of estimated FUV
intensities. The ratios involving CO(11--10) tend in general to indicate
somewhat higher densities of 10$^5$\,\cmcub\ or more.

Among the positions which were modeled, the position at
($-$13\arcsec,$-$79\arcsec) corresponds to the Southern Ridge (SR) that has
been modeled using PDR models by
\citet{Burton90b,Draine96,Draine00,Steiman97,Kaufman06,Sheffer11}.  Most of
these analyses used velocity-unresolved spectra of \CII, \OI, H$_2$ and
velocity-resolved CO spectra.  \citet{Burton90b} found that  the
observational data they had at the time agreed well with $G_0$ = 10$^4$ and
$n$ = $ 10^4$\,\cmcub. Their modeling does reproduce our observed \CII\
intensities extremely well, but overestimate the CO(7--6) intensity by
about a factor of five.  \citet{Steiman97} argued that the \CII\ emission
is likely to be dominated by emission from lower density gas surrounding
the PDR based on the clumpiness seen in the H$_2$ maps. \citet{Steiman97}
therefore separated the \CII\ emission into two components: a dense
component ($n\sim 10^5$\,\cmcub) filling 10\% of the beam, while a lower
density ($n\sim 10^3$\,\cmcub) fills the rest of the beam. A more refined
analysis using a merged PDR/\HII\ region model predicted that the denser
component corresponds to $G_0$ = 10$^4$ and $n$ = 5$\times 10^4$\,\cmcub\
\citep{Kaufman06}. Our velocity resolved \CII\ observations do not agree
with this picture. At the position of the SR there are  two velocity
components in \CII\ (Table~\ref{tab_n2023pos1}, Fig.\,\ref{fig_selspec}).
The red-shifted \CII\ component (V$_{lsr}$ = 10.7 \kms{}) originates in the
dense PDR, see Section~\ref{sect:selpos}, and most of the emission from the
blue-shifted \CII\ component originates in the PDR on the front side of the
C {\sc II} region. Our velocity resolved \CII\ observations therefore
suggest that only $\lesssim$ 20\% of the \CII\ emission is likely to come
from low density gas inside the nebula. The densities obtained here are
between 10$^4$ and 10$^5$\,\cmcub\ for the PDR emission from the SR
depending on which transition of CO {\em viz.,} (6--5) or (11--10) that is
used.

\section{Discussion}

NGC\,2023 has served as a testbed for PDR modeling in a number of
studies. The first PDR modeling was done by \citet{Black87} using the near
IR observations of fluorescent vibrationally excited H$_2$ emission in the
2 $\mu$m window by \citet{Hasegawa87} at the position 80\arcsec\ south of
HD\,37903. They found that to reproduce the observed H$_2$ emission line
they required a density of 10$^4$ cm$^{-3}$, a gas temperature, T$_k$
$\sim$ 85 K,  and a radiation field  G$_0$ = 300 times that of the Galactic
background star light. Since then the models as well as observational
constraints have vastly improved. Models utilizing near-IR fluorescent
H$_2$ lines at the SR (which includes the position 0",$-$80\arcsec{}), find
gas temperatures of the H$_2$ emitting region of 700 -- 1000 K, densities
of $\sim$ 10$^5$\,\cmcub, and a radiation field  $G_0 \sim$ 8000
\citep{Draine96,Draine00}.

The advantage of using fluorescent H$_2$ lines to probe the PDR
emission is that there are a large number of H$_2$ lines, see e.g.
\citet{Burton92,Burton98,McCartney99,Takami00}. The ratios of these lines
constrain the physical conditions, as well as the incident FUV field rather
well. The line ratios also discriminate between shock or fluorescent
excitation of the line emission, although there may be cases where one can
see both fluorescent and shock excited H$_2$. The  fluorescent H$_2$ lines
are only emitted from the hot gas in the PDR or in shocked gas and have no
contamination from low density gas inside the C {\sc II} region or from the
surrounding molecular cloud. However, they only probe the outer hot layers
of a PDR interface, and other tools are therefore needed to probe the
structure of the PDR. The general morphology of a PDR has been described in
many papers \citep[see e.g.,][]{tielens1985,Hollenbach91,Kaufman06}. A PDR
illuminated by an early B-star has a hot surface layer of $A_V$ $\lesssim$
1 -- 2, with atomic H and O, as well as C$^+$; a transition to H$_2$ and C
for $A_V \gtrsim$ 1 --2, where carbon starts to transition into a mixture
of C and CO; and for $A_V  \gtrsim$ 4 -- 6 most of the carbon is in CO,
while the oxygen not tied up in CO remains atomic to $A_V  \sim$ 10. Here
the PDR transitions into the cold molecular cloud (10 -- 40 K), which is
probed by CO and other traditional molecular gas tracers. The gas in the
PDR is heated by photoelectric heating from small dust grains and \CII\ is
excited by collisions with  hot electrons and hydrogen. The dominant
cooling lines are  the \CII\ 158 $\mu$m line and the \OI\ line at 63
$\mu$m.

The hot (1200 -- 200 K) intermediate layers are probed by \CII,  [O
{\sc I}], and  [C {\sc I}]; fluorescent  and pure rotational H$_2$ lines
and high-$J$ CO transitions, while the interior layer of a PDR is mostly
seen in pure rotational H$_2$ lines and high-$J$ CO transitions. This paper
focuses on velocity resolved spectroscopy of  the \CII\ 158 $\mu$m line and
high-$J$ CO lines observed with SOFIA and APEX. The high-$J$ CO lines,
especially CO(11--10) identify the dense PDR regions in the southeastern
part of the C {\sc II} region and show that CO resides in a hot ($\sim$ 100
K), thin layer surrounding the \CII\ emission. The close correspondence in
velocity (a few tenths of \kms{}) between \CII\ and CO(11--10) and/or
CO(7---6) indicate that most of the \CII\ emission originates in the dense
PDR  with only a minor part of the emission originating in low density gas
inside the C {\sc II} region, except for some regions in the dense
southeastern molecular ridge, where there is relatively strong blue-shifted
\CII\ emission without any counterpart in CO. The general agreement between
PDR emission and \CII\ emission is qualitatively seen in our integrated
intensity map of \CII\  overlaid on other PDR tracers
(Fig.\,\ref{fig_intplots}), which shows that the \CII\ emission traces the
boundary of the C {\sc II} region (limb-brightened) and the strongest \CII\
emission coincides with bright PDR filaments. Our PDR modeling (Section
\ref{sect:PDRmodeling}, see also Table~\ref{tab_pdrout}), also shows that
the  observed \CII\ emission can be explained with the \CII\ emission
coming from PDR regions with densities ranging from $\sim$ 1 $\times$
10$^4$ \cmcub\ -- 4 $\times$ 10$^5$ \cmcub, although the radiation field is
not very well constrained. There are, however, positions in the dense
southeastern ridge, where the lines are heavily blended and the \CII\
emission cannot be accurately identified as originating in  the hot PDR
shell. The \CII\ peak (30\arcsec,15\arcsec{}) is such a case. Here the
CO(11--10) emission profile is asymmetric with the dominant peak at 10.8
\kms, while the strongest \CII\ emission (peak temperature $\sim$ 74 K) is
centered at 9.1 \kms, where hardly any CO(11--10) emission is seen. At this
position the CO(7--6) line profile is symmetric and probably heavily
contaminated by the surrounding colder molecular cloud.  Position
(45\arcsec, $-$45\arcsec{}) is somewhat similar. At these positions it is
quite possible that more than 50\% of the \CII\ emission comes from low
density gas inside the C {\sc II} region. Here it would be useful to  have
observations of CO(8--7) (which can be observed from the ground, but only
under very good observing conditions), or CO(12--11) and CO(13--12), the
latter two can be observed with SOFIA and will have no contamination at all
from the surrounding cloud. However, an even better way to determine what
fraction of the ionized gas comes from the dense PDR vs. lower density gas
inside the C {\sc II} region is to do velocity resolved observations of the
\OI\ 63 $\mu$m line, which now can be done with the GREAT H-channel. The
\OI\ 63 $\mu$m line has a much higher critical density (4.7 $\times$
10$^5$\cmcub{}) compared to \CII\ (2.8 $\times$ 10$^3$ \cmcub{})
\citep{Meixner93}. If the \CII\ emission is dominated by gas with densities
 $<$ 10$^4$ \cmcub, it will not be seen in \OI. 

\section{Summary and Conclusions}

The NGC\,2023 reflection nebula was mapped in \CII\ and CO(11--10) with
the heterodyne receiver GREAT on SOFIA and  slightly smaller maps in
$^{13}$CO(3--2), CO(3--2),  CO(4--3), CO(6--5), and CO(7--6) were obtained
with APEX in Chile. The \CII\ emission was found to trace an expanding
ellipsoidal shell-like region  surrounded by a hot molecular shell seen in
high-$J$ CO lines. The expansion velocity of the PDR shell measured from
the  high-$J$ CO lines is $\sim$ 0.5 kms\ in the southeast and $\sim$ 1
\kms\ in the northwest , which is consistent with the size of the C {\sc
II} region. The \CII\ emission is dominated by dense (n = 10$^4$ -- 10$^5$
\cmcub{}) gas in the PDR interface between the C {\sc II} region and the
surrounding cold molecular cloud, except perhaps in  the southeast, where 
there may be a significant contribution from lower density
\CII\ emission inside  the C {\sc II} region. In the northwest, where the
surrounding molecular cloud is less dense, the densities  inside the C {\sc
II} region appear to be too low to excite \CII, and all the \CII\ emission
comes from the PDR shell surrounding the C {\sc II} region. Based on the
strength of the \thCII\ F=2--1 line, the \CII\ line appears to be somewhat
optically thick in the PDR shell with an optical depth of a few.  The
high-$J$ CO lines are very narrow (0.5 -- 1 \kms{}) and strong, indicating
that the CO emission in the PDR shell comes from a thin, hot, molecular
shell surrounding the \CII\ emission. The temperature of the CO emitting
PDR shell is $\sim$ 90 -- 120 K, with densities of 10$^5$ -- 10$^6$ \cmcub,
as deduced from RADEX modeling. The excitation temperature for \CII\  in
the PDR is similar to the CO in the hot molecular shell, or perhaps
slightly warner. The \CII\ lines are broader ($\sim$ 2 -- 3 \kms{}),
indicating that the \CII\ emitting region is more turbulent. PDR modeling
indicates that the densities in the \CII\ emitting region are also somewhat
lower, 10$^4$ \cmcub\ to a few times 10$^5$ \cmcub\ which is to be
expected, since the overpressure in the PDR forces the ionized gas to
expand into the \CII\ region. The \CII\ luminosity was found to be 1.6\% of
the total FIR luminosity for the whole reflection nebula. In the the dense
molecular ridge the fractional luminosity of \CII\ is lower, 1.1\% while
fractional luminosity of CO may be as high as 0.3\%.

Our high spatial and spectral imaging of \CII\  and high-$J$ CO lines
like CO(11--10) and CO(7--6)  gives us a unique insight to the morphology,
kinematics and physical conditions of the C {\sc II} region surrounding
NGC\, 2023. The only thing missing is additional high-$J$ CO lines like
CO(12--11) and CO(13--12), which are unaffected by the colder surrounding
molecular cloud. These lines, together with already existing \CII\ and
CO(11--10) data would help us further constrain the radiation field and
physical conditions in the PDR shell. Equally useful would be velocity
resolved spectra of the \OI\ 63 $\mu$m line, which can be used to
disentangle the fraction of diffuse and dense gas in the C {\sc II} region.
All these observations can be done with GREAT on SOFIA.

{}

\acknowledgement {This research has made use of data from the {\it Herschel} Gould Belt
survey (HGBS) project (http://gouldbelt-herschel.cea.fr). The HGBS is a
{\it Herschel} Key Programme jointly carried out by SPIRE Specialist
Astronomy Group 3 (SAG 3), scientists of several institutes in the PACS
Consortium (CEA Saclay, INAF-IFSI Rome and INAF-Arcetri, KU Leuven, MPIA
Heidelberg), and scientists of the Herschel Science Center (HSC). We thank
Dr. Michael Kaufman for valuable discussions about the physics and modeling
of PDRs. We thank the referee for valuable criticism enabling us to improve
the paper. A special thanks goes to Bill Vacca, who is a walking
encyclopedia and can answer any questions or solve any problems that we
have had. If he can't, he'll have the answer by the next day.}

\newpage
\begin{appendix}

\section{Comments on the selected positions} 
\label{sect:selpos}

\noindent
{\bf (0\arcsec, 0\arcsec{})} The position of HD\,37903. Probably a blend of
two PDRs, although they are too close together in velocity  to
 separated. South of HD\,37903 the high-$J$ CO lines split into
two components, but at the position of the star they are close to the cloud
velocity. \CII\ is much broader and it was with two velocity
components. \CII\ also shows a red-shifted wing. This red-shifted emission
extends toward SW and is roughly coincident with the somewhat blue-shifted
ridge of CO emission curving out toward HD\,37903. \\

 \noindent 
 {\bf (30\arcsec,+15\arcsec{})}  This is the position where
 \CII\ peaks, T$_{mb}$ $\sim$ 80 K. It is close to and probably associated
 with the bright fluorescent H$_2$ ridge \citet{Field98} call the Seahorse.
 Two velocity components were fitted to all observed lines, one close to
 the systemic velocity of the molecular cloud, 10.5 \kms, and one which
 is blue-shifted, 9.1 -- 9.4 \kms\ ( Table~\ref{tab_n2023pos1}). \\

\noindent 
{\bf ($-$13\arcsec,$-$79\arcsec{})}  This position is centered on the
southern ridge (SR), one of the most extensively modeled PDRs in NGC\,2023.
It is, however, a difficult position to model, because it coincides with
the blue outflow lobe from Sellgren C (Sandell et al. 2015, in prep.) and
the \CII\ line and high-$J$ CO lines show two velocity components: one
blue-shifted, and one red-shifted. The SR is definitely associated with the
red-shifted velocity component (V$_{lsr}$ $\sim$ 10.7 \kms{}), because
north-south cuts through the SR show that this velocity component coincides
with the SR. The SR also stands out in the high-$J$ CO channel maps, most
prominently in the channel centered on 11 \kms, see
Fig.\,\ref{fig_co11chanmap}, \ref{fig_co65chanmap}, \&
\ref{fig_co76chanmap}. It is still seen even in CO(4--3), but not as
clearly (Fig.\,\ref{fig_co43chanmap}). The good agreement in velocity
between \CII\ and CO(11--10) emission suggest that the \CII\ emission
originates from the dense PDR. The blue-shifted emission component has a
systemic velocity of  10 \kms, and almost certainly corresponds to a PDR on
the front side of the C {\sc II} region. This PDR is less dense than the
SR, and not detected in CO(11--10). Although it is clearly seen in CO(7--6)
and CO(6--5) it is hard to reliably estimate how much of the emission is
associated with the PDR, since both lines are affected by the  blue outflow
lobe from Sellgren C. The \CII\ line is more blue-shifted than the CO lines
and may include low density gas from the C {\sc II} region. Therefore only
the PDR component directly associated with the PDR is modeled. It is likely
that  the intensity of the high-$J$ CO transitions is somewhat
underestimated  since our observations are centered on the ridge and the
emission only fills about half the beam. However, since our PDR modeling
only use line ratios, the error from ignoring beam filling is relatively
minor.\\

\noindent 
{\bf (53\arcsec,$-$13\arcsec{}) \& (23\arcsec,$-$60\arcsec{})}  The
position (53\arcsec, $-$13\arcsec{}) coincides with the peak of CO(11--10)
emission peak as well as with the \crrl\ clump no. 3 \citep{Wyrowski00},
while (53\arcsec, $-$13\arcsec{}) is approximately centered on the  \crrl\
clump no.  2. No attempt was made to model the  \crrl\ clump no.1, because
the CO emission is contaminated by the strong red-shifted outflow from
Sellgren D. For both positions strong \CII\ emission is seen at about the
same velocity as \crrl, but at  (23\arcsec, $-$60\arcsec{}) the \CII\ line
is much broader, 2.5 \kms\ vs. 0.8 \kms\ for the \crrl\ emission line,
probably because  more of the extended emission associated with the ridge
is included and because \CII\ is not very sensitive to the density of the
emitting gas. A relatively strong blue-shifted \CII\ emission feature is
also seen at both positions without any counterpart in high-$J$ CO
emission. \\

\noindent
{\bf (45\arcsec,$-$45\arcsec{}) \& (60\arcsec,$-$60\arcsec{})} These
two positions probe the strong PDR ridge southeast of HD\,37903. A ``long''
integration spectrum  in \CII\ and CO(11--10) was obtained on (45\arcsec,
$-$45\arcsec{}), which is close to the peak position in CO(11--10). The
position (60\arcsec, $-$60\arcsec{}) coincides with the peak in low-$J$ CO
lines. Both positions show emission from two blended PDR components in
\CII\ (Fig.\,\ref{fig_selspec}, Table~\ref{tab_n2023pos1}). At  (45\arcsec,
$-$45\arcsec{}) the red-shifted component dominates the emission in both
\CII\ and CO(11--10). The second component is close to the cloud velocity.
Here \CII\ shows a strong blue-shifted wing making it difficult to separate
the emission from the PDR from the low density \CII\ inside the C {\sc II}
region, although PDR modeling was done for both components. At 
(60\arcsec,$-$60\arcsec{}) our high-$J$ CO spectra only show emission from
the PDR on the backside of the C {\sc II} region. The blue-shifted \CII\
component is almost certainly dominated by lower density gas inside the
nebula. Here only the red-shifted PDR component is modeled. The 
(60\arcsec,$-$60\arcsec{}) position was also observed by \citet{Jaffe90} in
CO(7--6) with a 34\arcsec{}-beam. Their results agree well with our 
findings.\\

\noindent
{\bf ($-$50\arcsec,+53\arcsec{})} This position is close to the
symmetry axis in the northwestern part of the nebula. All lines, except
\thCO{}(3--2) show double split lines. This is not due to self-absorption,
since there is no foreground gas that could absorb the CO(7--6) and
CO(6--5) emission. At CO(4--3) the emission from the backside of the nebula
is probably affected by self-absorption near the cloud velocity, making the
emission somewhat more red-shifted than the higher $J$ transitions. The PDR
modeling therefore only uses the CO(7--6) and CO(6--5) transitions and 
both PDRs are modeled separately.\\

\noindent 
{\bf (+17\arcsec, +113\arcsec{})} This position coincides with the
northern \thCO{}(3--2) peak and close to the crossing H$_2$ filaments,
which \citet{Field98} called the ``triangle'' (located at 16\arcsec,
+129\arcsec{}).  Here all the lines show only one velocity component at the
cloud velocity (Table~\ref{tab_n2023pos1}). The triangle was observed by
\citet{Takami00} in several H$_2$ emission lines in the near-IR with a
Fabry-Perot spectrometer providing spectral resolution, R, of 500 -- 2200.
This position is also close to the position  (+33\arcsec, +105\arcsec{})
observed by \citep{Burton98}, who obtained spectra from 1-- 2.5 $\mu$m, at
230 -- 430 spectral resolution.   \\

\section{CO channel maps}

\begin{figure}[h]
\begin{center}
\includegraphics[width=0.48\textwidth]{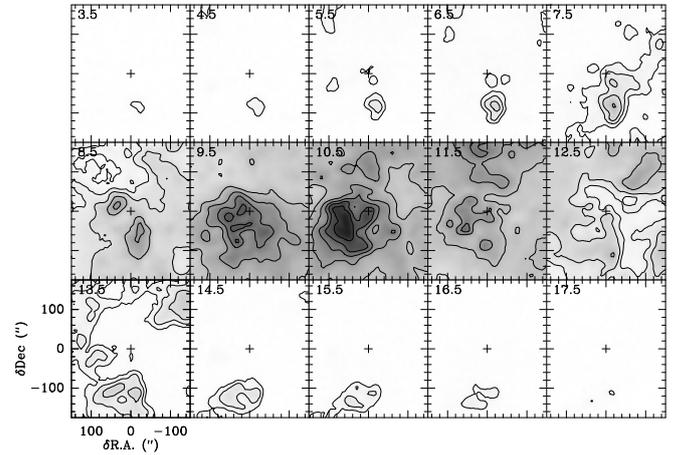}
\caption{Velocity-channel map for CO(3--2) with contours at 2, 5,
10 to 80\,K in steps of 10\,K. The `+' shows the position of HD\,37903.
\label{fig_co32chanmap}}
\end{center}
\end{figure}

\begin{figure}[h]
\begin{center}

\includegraphics[width=0.48\textwidth]{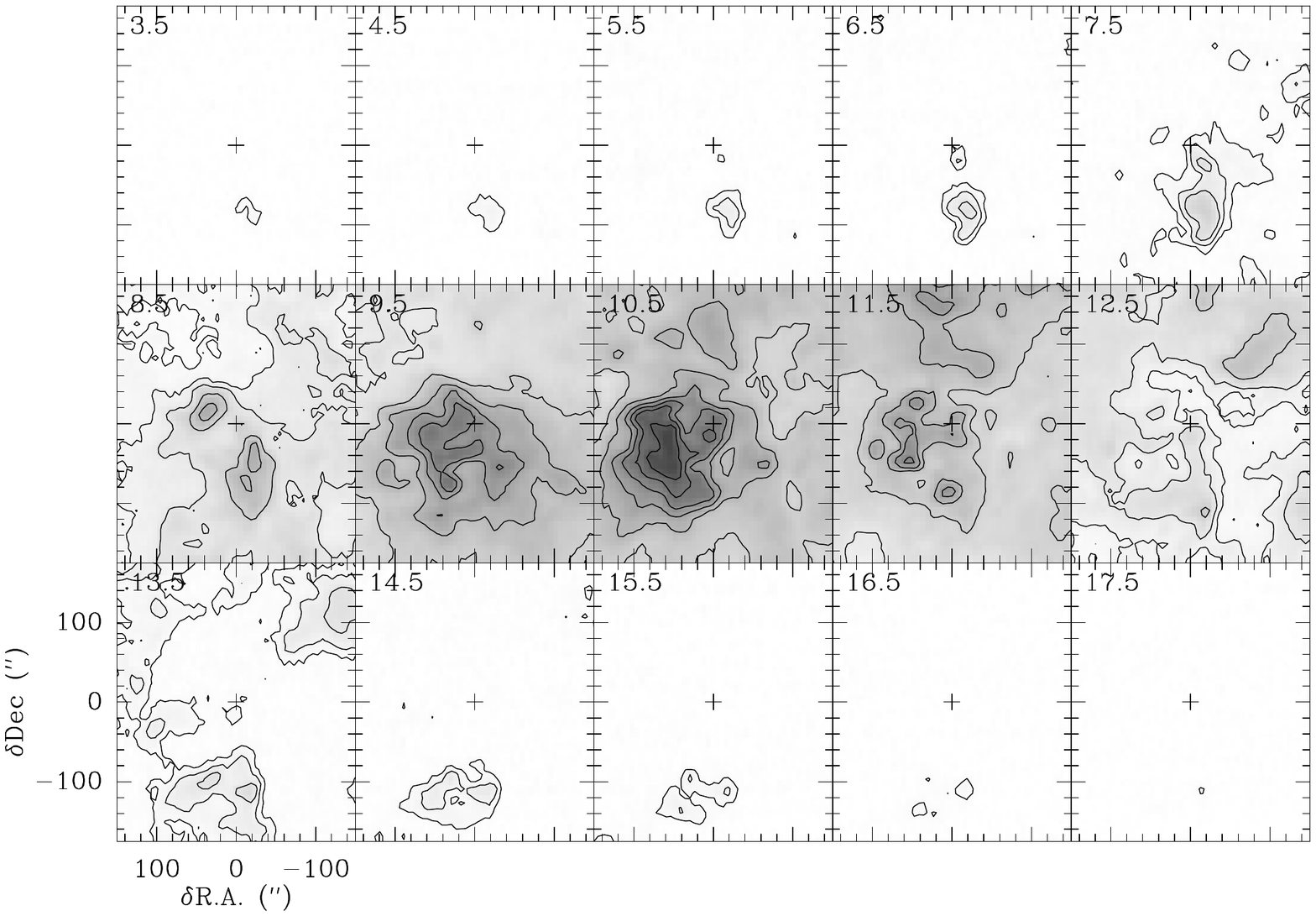}
\caption{Velocity-channel map for CO(4--3) with contours at 2, 5,
10 to 110\,K in steps of 10\,K. The `+' shows the position of HD\,37903.
\label{fig_co43chanmap}}
\end{center}
\end{figure}

\begin{figure}[h]
\begin{center}
\includegraphics[width=0.48\textwidth]{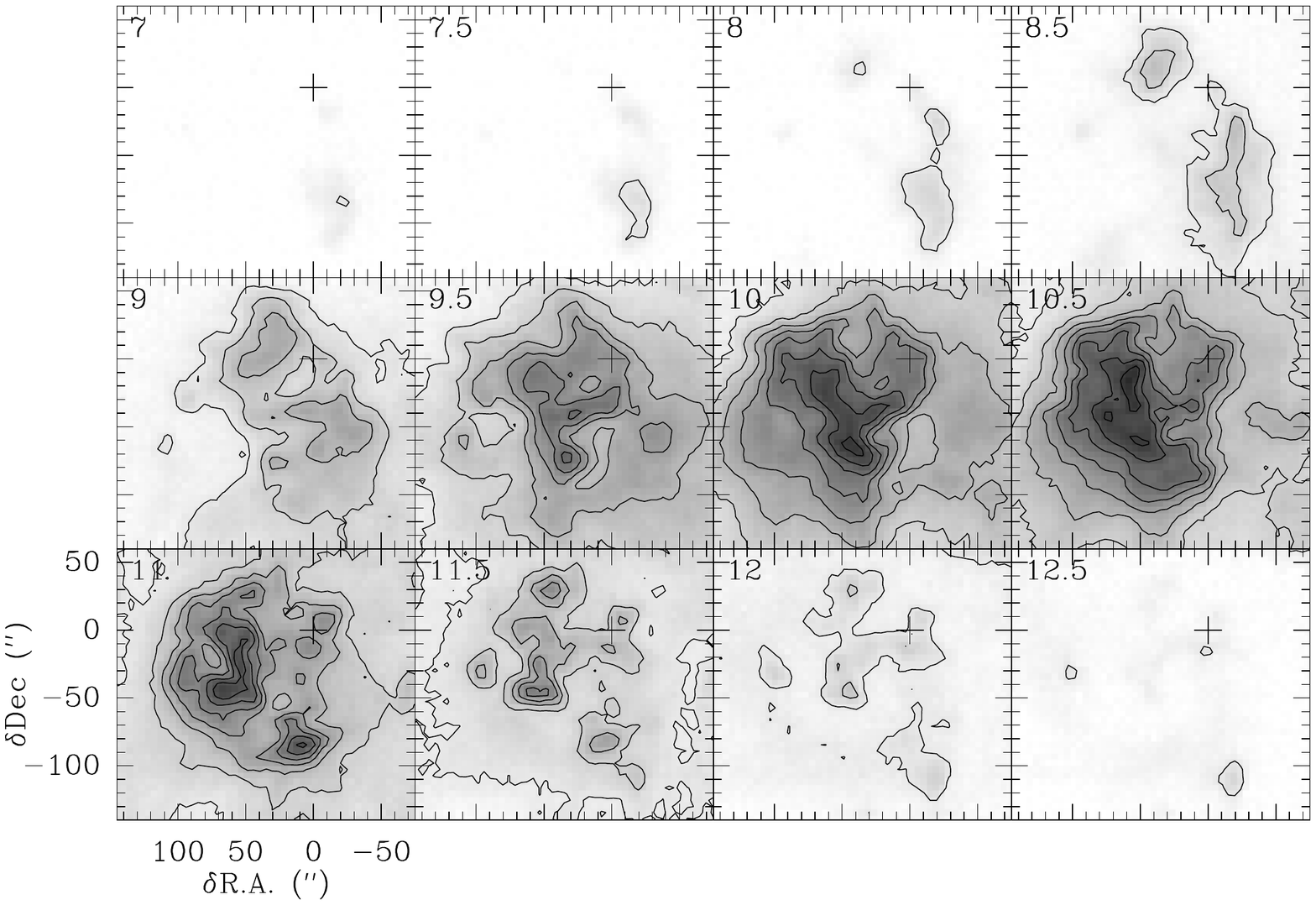}
\caption{Velocity-channel map for CO(6--5) showing the southeastern and
central part of the map with full resolution. The contour levels start  at
10 K and go to 100\,K in steps of 10\,K. The `+' shows the position of
HD\,37903. 
\label{fig_co65chanmap}}
\end{center}
\end{figure}

\begin{figure}[h]
\begin{center}
\includegraphics[width=0.48\textwidth]{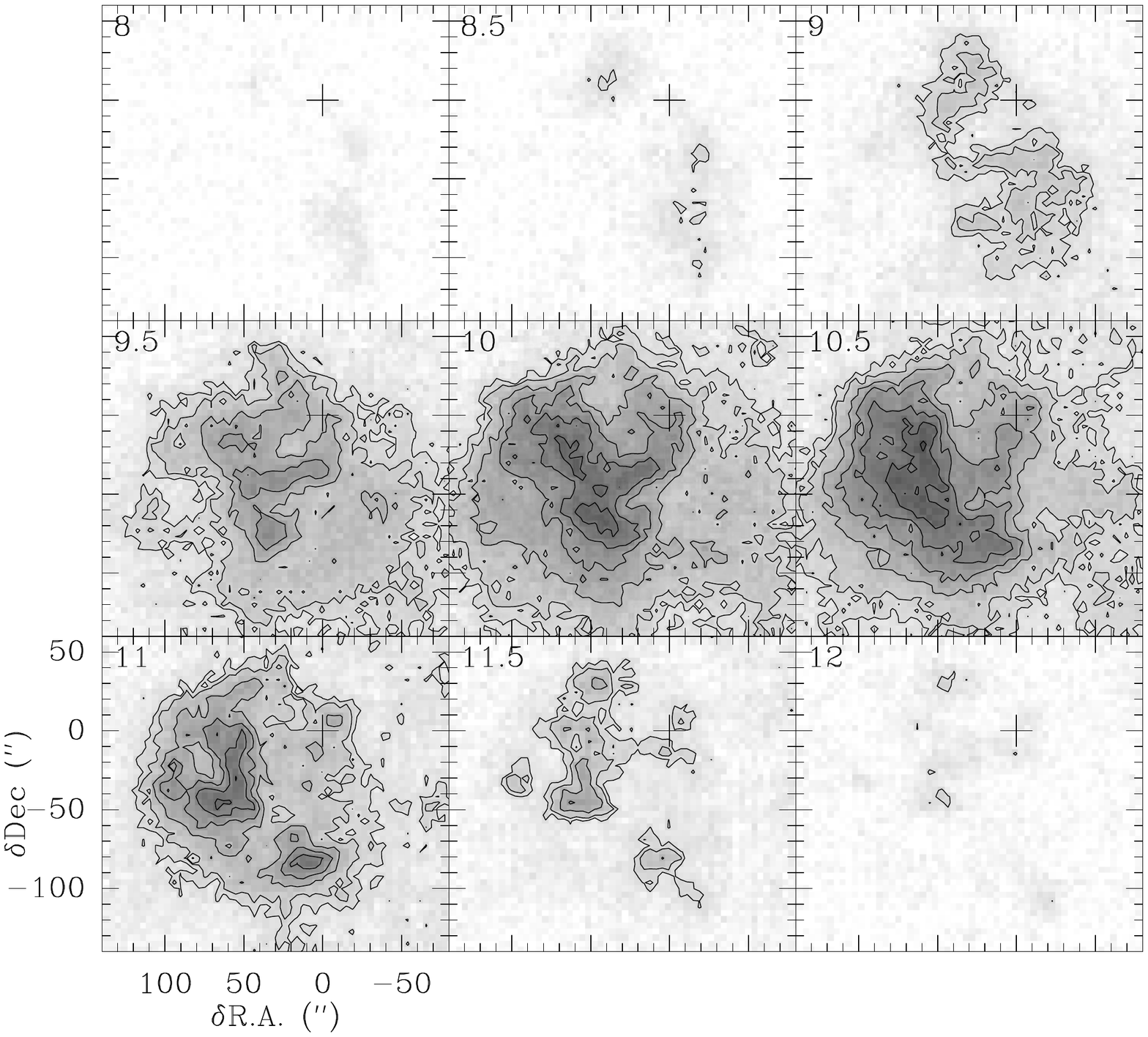}
\caption{Velocity-channel map for CO(7--6) showing the southeastern and
central part of the map with full resolution.  The contour levels are at
15\,K, and from 20\,K  to 100\,K in steps of 10\,K. The `+' shows the
position of HD\,37903.
\label{fig_co76chanmap}}
\end{center}
\end{figure}

\begin{figure}[h]
\begin{center}
\includegraphics[width=0.48\textwidth]{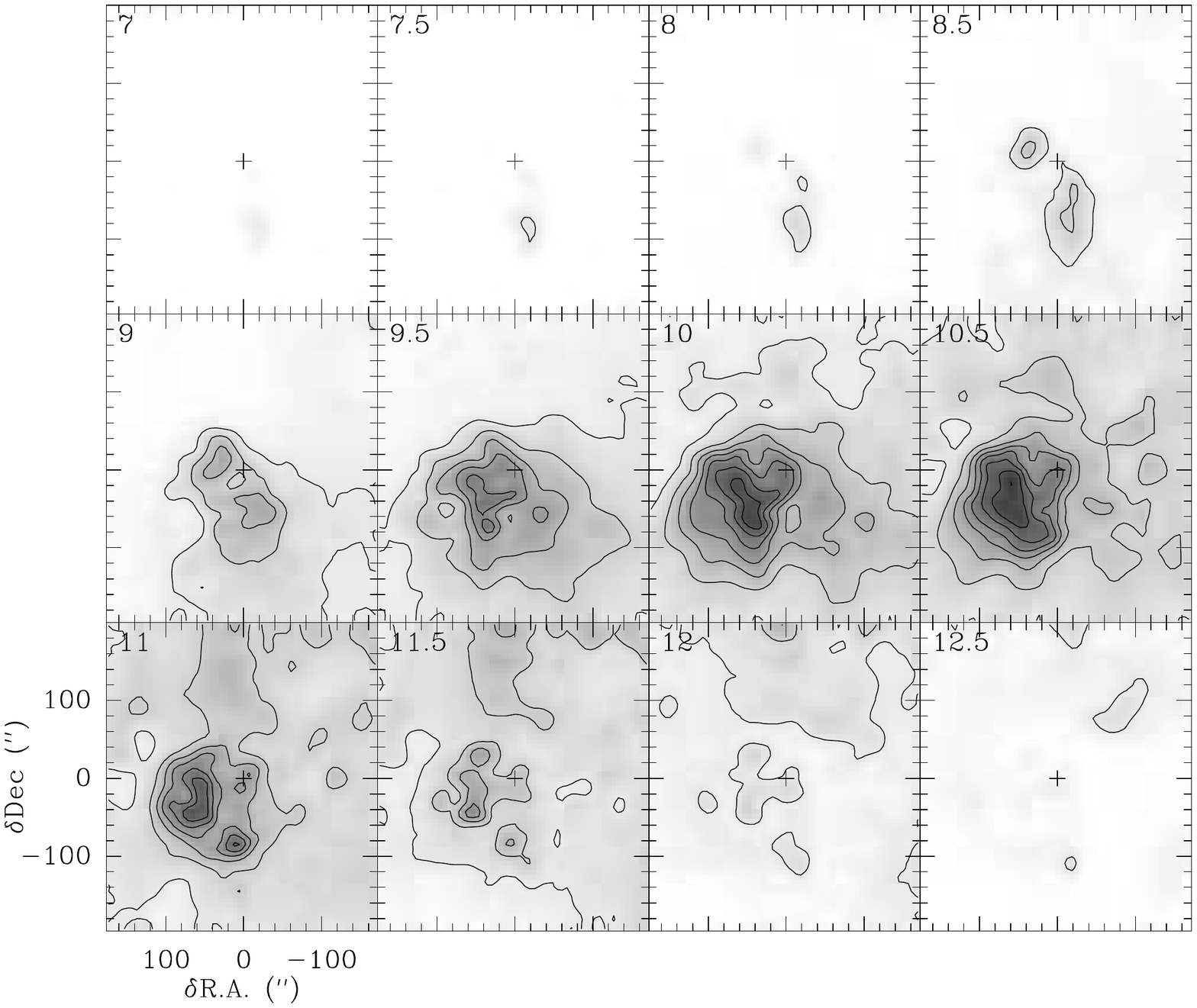}
\caption{Velocity-channel map for CO(6--5) smoothed to 16\farcs1
resolution showing the whole area that was  mapped. The contour levels go from
8\,K to 80\,K in steps of 8\,K. The `+' shows the position of HD\,37903.
\label{fig_co65CIIchanmap}}
\end{center}
\end{figure}

\begin{figure}[h]
\begin{center}
\includegraphics[width=0.48\textwidth]{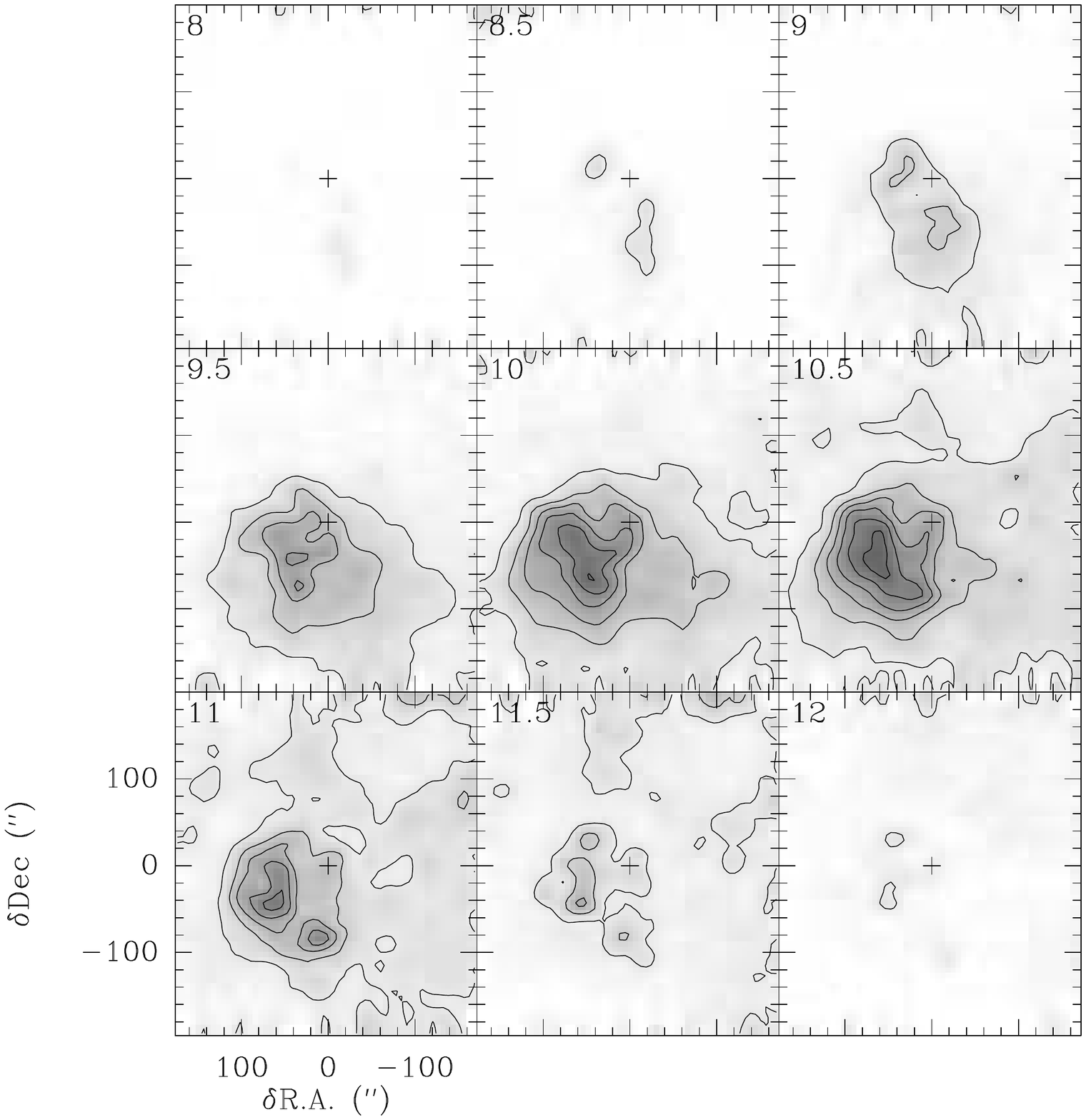}
\caption{Velocity-channel map for CO(7--6) smoothed to 16\farcs1
resolution showing the whole area that was mapped. The contour levels go from
6.3\,K  to 63\,K in steps of 6.3\,K. The `+' shows the position of
HD\,37903.
\label{fig_co76CIIchanmap}}
\end{center}
\end{figure}

Figures\,\ref{fig_co32chanmap} to \ref{fig_co65chanmap} show the
channel maps of CO(3--2), CO(4--3), CO(6--5), and CO(7--6). The fields of
view of the CO(6--5) (Fig.\,\ref{fig_co65chanmap}) and CO(7--6)
(Fig.\,\ref{fig_co76chanmap}) channel maps were restricted to the
southeastern and central part of the nebula, in order to visualize the
strong CO emission from the PDR better. Figures\,\ref{fig_co65CIIchanmap}
and  \ref{fig_co76CIIchanmap} show the channel maps for CO(6--5) and
CO(7--6) but smoothed to the 16\farcs1 resolution of the \CII\ data in
order to capture the fainter emission in the northwestern part of the map
better.

\end{appendix}
\end{document}